\documentclass[12pt,notitlepage]{article}

\usepackage{amsfonts}
\usepackage{amssymb}
\usepackage{standalone}
\usepackage{graphicx}
\usepackage{amsmath}
\usepackage{amsthm}
\usepackage{endnotes}
\usepackage{setspace}
\usepackage{setspace}
\usepackage{natbib}
\usepackage{xcolor}
\usepackage{bigints}
\usepackage{tikz,pgfplots}
\pgfplotsset{compat=1.17}
\usepackage[title]{appendix}
\usepackage{bbm}
\usepackage{enumerate}
\usepackage{enumitem}
\newtheorem{theorem}{Theorem}

\newtheorem{corollary}{Corollary}

\newtheorem{Lemma}{Lemma}

\newtheorem{Proposition}{Proposition}

\newcommand{\uS}{\hat{u}^{S}}

\newcommand{\cav}{\operatorname{cav}}
\newcommand{\cavk}{\operatorname{cav}_k}
\newcommand{\VS}{V}

\newcommand{\E}{\mathbb{E}}
\newcommand{\supp}{\operatorname{supp}}

\newcommand{\Ext}{\operatorname{Ext}}

\definecolor{cardinal}{rgb}{0.77, 0.12, 0.23}
\usepackage[colorlinks=true]{hyperref}
\hypersetup{
     colorlinks=true,
     linkcolor=cardinal,
     filecolor=cardinal,
     citecolor = black,
     urlcolor=cardinal,
     }
\makeatletter

  \usepackage[margin=.9in]{geometry}

\usepackage{chngcntr}
\counterwithout{figure}{section}
\counterwithout{table}{section}

\usepackage{subcaption}
\begin{document}

\title{Persuasion with Coarse Communication\thanks{We are grateful to  Avidit Acharya, Steven Callander, Oguzhan Celebi, Doruk Cetemen, Mine Su Erturk, Francoise Forges, Matthew Gentzkow, Arda Gitmez, Edoardo Grillo, Matthew Jackson, Semih Kara, Tarik Kara, Emin Karagozoglu, Silvana Krasteva,  Elliot Lipnowski, Ludvig Sinander, Cem Tutuncu, Robert Wilson, Kemal Yildiz, Weijie Zhong, seminar audiences at Stanford University, Bilkent University, Stony Brook Game Theory Festival 2022 \& 2023, and Econometric Society Meetings in 2020 \& 2021 for helpful comments.}}

\author{\href{https://www.yunusaybas.com/}{Yunus C. Aybas}\thanks{
Texas A\&M University, Department of Economics. Corresponding author: \href{mailto:aybas@tamu.edu}{aybas@tamu.edu}.} \and
\href{https://erayturkel.github.io}{Eray Turkel}\thanks{
Stanford University, Stanford Graduate School of Business.}
}

\maketitle
\begin{abstract}
In many expert–decision maker settings, information is richer than the language used to convey it. Motivated by this communication friction, we study Bayesian persuasion when the sender is constrained to use $k$ messages. We show that the sender's value is given by a $k$-point analogue of concavification, which we call $k$-concavification. An optimal information structure can be chosen with affinely independent posterior support, allowing the problem to be reduced to a lower-dimensional persuasion problem and then solved by standard concavification. We derive a tight bound on the value of communication capacity that applies to general persuasion games: the gain from a $(k+1)$\textsuperscript{st} message is at most $2/(k-1)$ times the value attainable with $k$ messages. Finally, we solve a class of belief-threshold games in which the receiver chooses between a safe default and several risky actions, the sender gets zero from the default and the same positive payoff from any risky action, and a risky action is taken only when the corresponding posterior probability exceeds a threshold. We characterize the optimal coarse information structure, derive comparative statics in the prior and the threshold, and extend the analysis to heterogeneous thresholds and heterogeneous sender values across risky actions.

\vspace{.3cm}

\vspace{.3cm}
		\end{abstract}

\thispagestyle{empty}
\newpage
 \section{Introduction}

In many expert--decision maker interactions, the relevant information is richer than the language available to report it. Credit rating agencies summarize complex issuer risk using a small number of rating categories. Regulators translate detailed inspections into hygiene scores or energy-efficiency labels. Schools compress months of performance into letter grades. In each case, a large set of states must be conveyed through a small set of admissible messages.

The canonical Bayesian persuasion model of \citet{KG11} stands in sharp contrast to these practical realities. It grants the sender a rich unconstrained message space. The sender can separate states perfectly or  recommend any action without restriction. But when communication is inherently coarse, as it so often is, such flexibility is no longer feasible.

We revisit Bayesian persuasion when this flexibility is absent. There is a sender and a receiver, the state is drawn from a finite set, and the sender commits ex ante to an experiment that maps states into at most $k$ messages. After observing the realized message, the receiver updates and chooses an action. Relative to the canonical persuasion problem, the only added restriction is that the distribution of posterior beliefs must have support of size at most $k$.

Our model captures settings in which the relevant language is determined by regulation, organizational protocol, platform design, or a shared institutional convention before the sender faces a particular persuasion problem. The sender can choose what to learn and how to map what she learns into the available messages, but she cannot add new messages or change the set of admissible categories.

An example we analyze is commercial lending. A loan officer can decide how thoroughly to investigate a borrower---by requesting financial statements, verifying collateral, conducting interviews, or following up on inconsistencies---and these choices determine the experiment. But once the investigation is complete, the officer must summarize her assessment using the bank's internal rating system, a fixed set of discrete grades \citep{TreacyCarey1998,FRBAdvancedApproaches217122}.\footnote{Internal bank rating systems use discrete grades rather than unrestricted narrative assessments. \citet{TreacyCarey1998} report that among the fifty largest U.S.\ banks, the number of grades ranges from two to the low twenties, with a median of five Pass grades. For banks subject to the advanced-approaches framework, the wholesale obligor rating system must have at least seven discrete rating grades for non-defaulted obligors \citep{FRBAdvancedApproaches217122}.} The same logic applies more broadly whenever standardized labels, scores, ratings, or grades are fixed to make complex assessments easier to interpret, compare, validate, and use.\footnote{Similar patterns also arise in other banking regulations. For instance, complex assessments of risk are translated into a credit rating and disclosure categories so that supervisors and market participants can compare positions across institutions and jurisdictions, and link those classifications to regulatory consequences such as capital requirements \citep{BCBSCapitalDisclosure2012}.}

The seemingly simple restriction on the signal space changes the geometry of persuasion. Bayes plausibility still requires the induced posteriors to average to the prior, but now the sender must support that distribution on at most $k$ points. Persuasion becomes a problem of allocating scarce communication capacity.  This raises several fundamental questions.  What is the sender's value with a limited number of messages? How does message scarcity affect the sender's payoff? What is the value of an additional message? What are the optimal coarse information structures?

The contribution of this paper is to characterize how exogenous constraints on message capacity impact information disclosure. We characterize the sender's value under an arbitrary message space, derive a sharp bound on the value of an additional message, and explicitly solve a tractable and realistic class of persuasion problems that makes the economics of message scarcity transparent.

\medskip
\noindent
\textbf{k-Concavification.}
Our first contribution is a general characterization of the sender's value. In the unconstrained persuasion problem, the sender's payoff is characterized by \textit{concavification}. We show that with $k$ messages, the relevant object is a $k$-point analogue of concavification, which we call \emph{$k$-concavification}. It gives the best Bayes plausible decomposition of the prior into at most $k$ posterior beliefs and therefore characterizes the sender's optimum for every prior and every message space. When $k$ is small, a large number of states must be bundled into a few messages, and as $k$ rises, the set of attainable posterior distributions expands, eventually converging to the rich-message benchmark.

\medskip
\noindent
\textbf{Compress-then-concavify.} Our second result shows that the support of an optimal information structure can be chosen to be \textit{affinely independent}. This observation yields a useful interpretation. One can view the sender as first compressing the state space into a $k$-element \emph{pseudo-state space}, and then solving a canonical persuasion problem on that compressed space. Thus, optimal coarse persuasion can be stated as an  \emph{optimal compression} problem. We show that this logic  extends beyond Bayesian persuasion to other belief-based environments with limited messages, including cheap talk with state-independent sender preferences.

\medskip
\noindent
\textbf{Value of an additional message.} A finite message space also raises a natural comparative-static question. In many applications, adding one more category is costly. New labels must be defined, standardized, monitored, and learned. Finer communication may also increase cognitive burden. We therefore study how much the sender can gain from an additional message.

Our main result provides a tight bound on the value of an additional message that applies to all persuasion environments. The bound states that granting the sender a $(k+1)$\textsuperscript{st} message can increase her payoff by at most $\frac{2}{k-1}$ times the $k$-message payoff. This bound reveals a fundamental property of coarse communication. The maximum percentage returns to message capacity are necessarily diminishing, and decline rapidly as $k$ grows.

The endogenous design of the message space is an interesting but separate problem. Adding categories may require costly coordination, standardization, verification, training, and compliance, and a finer language may impose additional cognitive and operational burdens on receivers. By taking coarseness as exogenous, the paper isolates the consequences of communicating through a fixed coarse message space. The value bounds below provide a benchmark for assessing when the benefits of a richer communication system justify the costs of creating it.

\medskip
\noindent
\textbf{Belief-threshold games.}
To sharpen these general insights, we study a tractable and widely applicable class of persuasion problems that we call \emph{belief-threshold games}. In this class, the receiver chooses between a safe default action and several risky actions. Risky action $i$ is optimal only when the posterior probability of state $i$ exceeds a fixed threshold. The sender's payoff is state independent, she receives zero from the default and the same positive payoff from any risky action. Her objective is therefore simply to move the receiver away from the default, and the threshold is a measure of the difficulty of persuasion.

Within this class, we characterize the optimal information structure for any number of states and messages. Three principles govern the solution. First, if some messages have to induce the default action due to Bayes plausibility, the information structure eliminates slack at risky posteriors by setting the probability of the risky action inducing state exactly at the belief threshold. Second, it preserves the relative proportions of the remaining states, thereby freeing probability mass for the action-supporting states. Third, it prioritizes risky actions corresponding to the highest-probability states first.

These principles lead to a sharp characterization. If the cumulative prior probability of the $k$ most likely states is large enough, then the sender can induce $k$ risky actions and attain the full payoff. The optimal design separates the top $k$ states and pools each of them with the remaining states in prior proportions. If that cumulative mass is too small, then at least one message must induce the default action, so at most $k-1$ risky messages are feasible. In that case, the sender targets the $k-1$ most likely states, puts each targeted state exactly at the threshold in its corresponding posterior, assigns zero probability to the other targeted states there, and preserves the remaining tail in prior proportions. Overshooting the threshold or distorting the pooled tail uses scarce message capacity inefficiently and lowers the ex ante frequency of risky actions.

This characterization yields comparative statics for the value of communication capacity. Below the cutoff at which the default action can be eliminated, an additional message is used to target the next most likely state, so the marginal gain declines as messages are allocated to less likely states. At the cutoff, there can be a discrete \emph{last-mile} jump, because one more message eliminates the default altogether. Beyond that cutoff, further messages have no value. More concentrated priors raise the sender's value for every $k$, lower the number of messages needed to eliminate the default, and shrink the last-mile jump. Higher thresholds make persuasion uniformly harder, lowering value and shifting the cutoff to larger $k$.

We also extend the analysis to heterogeneous belief thresholds, a valued subset of risky actions, and heterogeneous sender values across risky actions.

\medskip
\noindent
\textbf{Relationship to the literature.} A broad literature studies how coarse communication shapes information transmission and decisions, including organizational economics \citep{arrow1974limited, cremer2007language,sobel2015broad}, matching \citep{mcafee2002coarse, hoppe2011coarse}, certification \citep{ostrovsky2010information, harbaugh2018coarse, zapechelnyuk2020optimal}, and bounded rationality \citep{rubinstein2000economics, wilson2014bounded}.

Within strategic communication, contributions address the use of vague language \citep{lipman2025language}, coarse communication \citep{jager2011voronoi,de2003game, blume2000coordination, blume2013language}, and coarse understanding \citep{hagenbach2020cheap}. Our work departs from this literature by focusing on environments with preference misalignment between the expert and the decision maker.

Our friction is also distinct from other limits on communication in persuasion. In models with exogenous noise \citep{blume2007noisy, akyol2016information} or information-theoretic costs of informativeness \citep{gentzkow2014costly, bloedel2018persuasion, wei2018persuasion}, the restriction applies to the informativeness of a given message. \citet{parakhonyak2026persuasion} study a different robustness concern: the sender evaluates persuasion mechanisms by minimax regret when she is uncertain about the state distribution, the receiver's prior, and the receiver's utility.\footnote{A related prior-free approach to comparing experiments is developed by \citet{rosenthal2026priorfree}, who rank experiments for a maxmin decision maker observing the distribution of signals generated by a known experiment under an unknown state distribution.} Here, any one message can be perfectly precise and can even reveal the state. The friction is instead one of dimensionality. This captures settings in which the bottleneck is a standardized menu of categories rather than noisy transmission, costs of precision, or ambiguity about the prior.

\citet{le2019persuasion} study repeated Bayesian persuasion when communication takes place over a channel with both noise and cardinality constraints. Their results characterize long-run sender utility in terms of the channel’s capacity. In contrast, we focus on one-shot persuasion games without noise or attention.

In concurrent work, \citet{lyu2023coarse}  study a linear persuasion problem with a finite message space.
They primarily focus on the case where the sender and receiver share aligned preferences and determine which part of the state space should be more finely partitioned. We consider their work to be a complement to ours, emphasizing different questions that arise with limited communication capacity.

\citet{curello2022comparative} also study linear persuasion and show how comparative statics over the sender’s value function extend to constrained message spaces.

In a framework that nests Bayesian persuasion (sender commitment) and mechanism design (receiver commitment), \cite{le2024mediated} study mediated communication with finitely many messages. They show that the $k$-concavification technique developed in this paper can be used to characterize payoffs in mediated communication.

\section{The Model}\label{setup}
\textbf{Primitives.}
We study Bayesian persuasion with a constraint on the sender’s message space. Let $\Omega$ be the finite state space and $A$ be the compact action space. The sender (she) and the receiver (he) share a prior $\mu_0\in\Delta(\Omega)$. The sender’s and receiver’s payoffs are $u^S,u^R:A\times\Omega\to\mathbb{R}$, respectively.

\medskip
\noindent
\textbf{Coarse communication.}
The sender communicates using a message space $M$ with cardinality $k$. Let  $ n:=\min\{|\Omega|,|A|\}$, and assume \(1 < k < n.\)  The case $k=1$ is excluded because a single message cannot convey information beyond the prior. The upper bound is without loss of generality for the reasons discussed in \cite{KG11}.\footnote{First, by Carath\'eodory’s theorem on the $(|\Omega|-1)$-dimensional simplex $\Delta(\Omega)$, any Bayes plausible distribution over posteriors can be supported on at most $|\Omega|$ points. Second, in the unconstrained persuasion problem, an optimal experiment induces at most $|A|$ posteriors. If $k\ge \min\{|\Omega|,|A|\}$, the cardinality constraint is slack.}

\medskip
\noindent
\textbf{Timing and strategies.}
The sender publicly commits to an experiment \( \pi:\Omega\to\Delta(M) \). We often denote $\pi$ as a collection of conditional probability mass functions $\{\pi( \cdot | \omega) \}_{\omega \in \Omega}$. Once $\pi$ is committed, a state $\omega\in \Omega$ is drawn according to $\mu_0$. Given the realized state $\omega$, the sender draws a message  according to the committed information structure $\pi( \cdot | \omega)$ and communicates the realized message $m$.

Observing $m$, the receiver’s posterior is given by Bayes’ rule, and denoted by $\mu(\omega | m)$.  The receiver with posterior $\mu\in\Delta(\Omega)$ chooses an action \(\hat a(\mu) \) that maximizes \(\sum_{\omega}\mu(\omega)u^R(a,\omega),\) with ties broken in favor of the sender.

\medskip
\noindent
\textbf{Sender's payoff.}
We define the sender's interim payoff over the posterior space by
\(
\uS(\mu) := \sum_{\omega }\mu(\omega)u^S \big(\hat a(\mu),\omega\big).
\)
Given $\pi$, the sender’s
ex-ante payoff can be calculated by taking expectation over $\hat{u}^S$ with respect to the ex-ante distribution of messages.

\medskip
\noindent
\textbf{Posterior distributions and feasibility.} It is often convenient to pass to the induced distribution of posteriors \( \tau \left(\mu(\cdot\mid m)\right) =  \sum_{\omega}\mu_0(\omega)\pi(m | \omega).\) We refer to $\tau$ as an \textit{information structure}.

We say that $\tau$ is \textit{Bayes plausible} if $\E_{\tau}[\mu]=\mu_0$ and its support has at most $k$ elements, i.e. $|\supp(\tau)|\le k$. We define the set of Bayes plausible information structures as $\mathcal{I}(k,\mu_0)$.

As in \cite{KG11}, every $\pi$ with at most $k$ messages induces some $\tau \in\mathcal I(k,\mu_0)$, and every $\tau\in\mathcal I(k,\mu_0)$ can be implemented by an experiment $\pi$ with at most $k$ messages. Hence the sender’s problem can be stated as maximizing the ex-ante expected payoff $\E_{\tau}\big[\uS(\mu)\big]$ with the constraint that $\tau\in\mathcal I(k,\mu_0)$.

A maximizer exists. We omit the proof and provide a short sketch. Parameterize posterior distributions by $(\tau,\mu_1,\ldots,\mu_k)\in\Delta(\{1,\dots,k\}) \times\Delta(\Omega)^k$ subject to $\sum_{i=1}^k \tau_i\mu_i=\mu_0$. The feasible set is a closed subset of the compact set $\Delta(\{1,\dots,k\}) \times\Delta(\Omega)^k$. Hence, it is compact. The objective $(\tau,\mu_1,\ldots,\mu_k)\mapsto \sum_{i=1}^k \tau_i\uS(\mu_i)$ is upper semicontinuous. By the extreme value theorem, a maximizer exists.

Finally, we denote the sender’s value with \(k\) messages, $\max_{\tau\in\mathcal I(k,\mu_0)} \E_{\tau}\big[\uS(\mu)\big]$, as \( \VS(k,\mu_0).\)

\section{Geometric Solutions to Coarse Persuasion Games}

\textbf{k-Concavification.} Let \(\cav(\hat u^S)\) denote the convex hull of the graph of sender’s interim payoff \(\hat u^S\). \cite{KG11} shows that if $k \geq n$  then whenever \((\mu_0,z)\in \cav(\hat u^S)\)  there exists  a Bayes-plausible
posterior distribution that achieves the ex-ante payoff \(z\) under the prior $\mu_0$.  Related geometric characterizations also appear in cheap talk with state-independent sender preferences and in verifiable-information settings. Existing tools, however, make the same rich message space assumption. In this section, we show that the insights gained from existing results can be effectively extended to coarse communication.

Under coarse communication, \((\mu_0,z)\in \cav(\hat u^S)\) need not be feasible when the associated posterior distribution requires more than \(k\) posteriors. To encode the cardinality constraint in taking convex combinations, we define the \textit{k-convex hull} of the graph of \(\hat u^S\):
\[
\cavk(\hat u^S)
:=\Bigl\{(\mu,z) \  \mid \  \exists (\tau_i,\mu_i)_{i=1}^k
\text{ s.t. }
\tau_i\ge 0,\ \sum_{i=1}^k \tau_i=1,\ \sum_{i=1}^k \tau_i\mu_i=\mu,
\text{ and }
z\le \sum_{i=1}^k \tau_i \hat u^S(\mu_i) \Bigr\}.
\]
We call the map \(\mu_0\mapsto \sup\{z:(\mu_0,z)\in \cavk(\hat u^S)\}\) the \emph{k-concavification}
of \(\hat u^S\). As \(k\) increases, \(\cav_k(\hat u^S)\) expands towards \(\cav(\hat u^S)\), recovering the
unconstrained benchmark at $k=n$.

The next result shows that if $(\mu_0, z) \in \cavk(\hat{u}^S)$, then there exists a feasible posterior distribution that yields the sender a payoff of $z$ when the prior is $\mu_0$, and vice versa. So,  the sender’s value with \(k\) messages equals the \(k\)-concavification value at \(\mu_0\).

\begin{Proposition}\label{theorem:Equivalence}  \(\tau\in\mathcal I(k,\mu_0)\) is optimal if and only if \( \E_{\tau}\big[\hat u^S(\mu)\big] = \sup\{z:(\mu_0,z)\in \cavk (\hat u^S)\}. \) \end{Proposition}

\begin{figure}[h!!]
  \centering
   \begin{subfigure}{0.31\textwidth}
  \includegraphics[width=\linewidth]{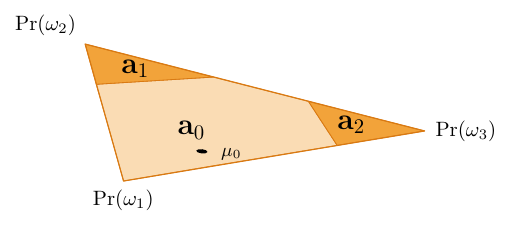}
  \caption{Receiver actions}
  \label{fig:demo_utility-receiver}
  \end{subfigure}\hfill
  \begin{subfigure}{0.31\textwidth}
  \includegraphics[width=\linewidth]{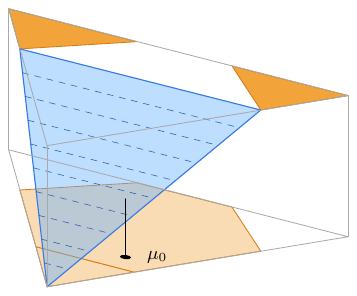}
  \caption{Optimal with $k=3$}
  \label{fig:demo_utility-3}
   \end{subfigure}\hfill
   \begin{subfigure}{0.31\textwidth}
  \includegraphics[width=\linewidth]
  {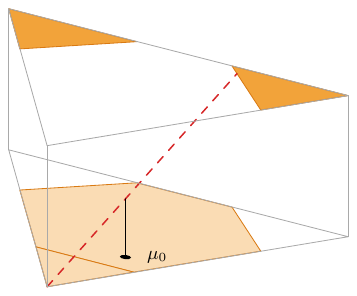}
  \caption{Optimal with $k=2$}
    \label{fig:demo_utility-2}
  \end{subfigure}
  \caption[]{Belief regions and optimal experiments in a three-state persuasion problem.}
  \label{fig:demo_utility}
\end{figure}

\textbf{A Geometric Illustration of the Result.}
To isolate the geometry of Proposition~\ref{theorem:Equivalence}, consider a three-state persuasion problem.
The receiver's best response partitions the belief simplex into three action regions.  The sender receives a high payoff in two of these regions and a low payoff in the remaining region. Her problem is therefore to choose at most \(k\) posterior beliefs whose weighted average is \(\mu_0\) and whose average payoff is as large as possible.  Figure~\ref{fig:demo_utility-receiver} displays this payoff geometry where the shaded regions identify the receiver's optimal action at each posterior belief and the black dot denotes the prior $\mu_0 \approx \left(\frac{73}{120}, \frac{14}{120}, \frac{33}{120}\right)$

Figures~\ref{fig:demo_utility-3} and~\ref{fig:demo_utility-2} illustrate the geometric consequences of coarseness. With three messages, the sender can induce three distinct posteriors: two in high-payoff action regions and one residual posterior in the low-payoff region required by Bayes plausibility. This separates the two favorable regions while keeping the posterior distribution feasible.

When constrained to two messages, the sender can induce at most two posteriors. One posterior can be placed in a high-payoff region, but the second must balance the prior and therefore falls in the low-payoff region. The loss from coarseness is thus a geometric consequence of having fewer posterior support points with which to span the prior.

\pagebreak

\begin{figure}[b!!]
  \centering
 \includegraphics[width=.82\textwidth]{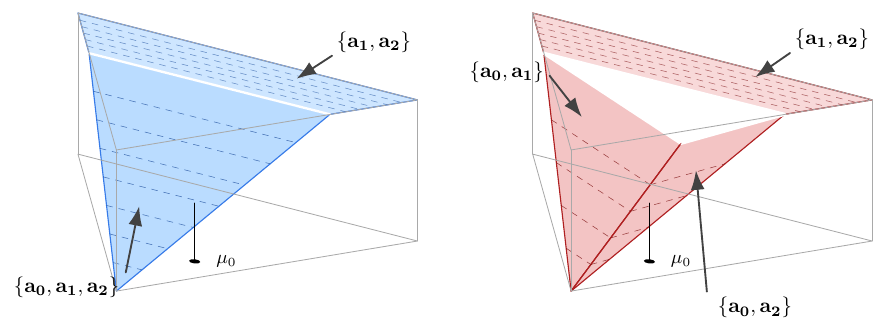}
            \caption{The highest achievable sender utility with 3 signals (left) and 2 signals (right). The black line corresponds to the prior. The set of receiver actions that are induced under the optimal information structures are denoted in curly brackets}
              \label{fig:demo_kcav}
\end{figure}

Figure~\ref{fig:demo_kcav} plots the resulting $k$-concavification surfaces. It compares the convex hull of $\hat u^S$, $\operatorname{cav}_3(\hat u^S)$ with the $2$-convex hull of $\hat u^S$, $\operatorname{cav}_2(\hat u^S)$. The gap between the two surfaces measures the marginal value of a third message. This value is strictly positive for interior priors, where the $k=2$ constraint binds and reduces the sender's expected payoff.

\noindent
\textbf{Example: Commercial Lending.}
The illustration above can also be interpreted as an application to commercial lending. The three states can be read as weak, stable cash-flow, and strong-growth borrower profiles, and the prior \(\mu_0\) as the bank's initial assessment of the borrower pool. The credit committee is the receiver. Its default action \(a_0\) is rejection, while \(a_1\) and \(a_2\) are approval products or internal grades suited to different favorable borrower profiles. The shaded regions in Figure~\ref{fig:demo_utility-receiver} therefore describe the posterior beliefs under which the committee is willing to move away from rejection and choose each approval option.

The loan officer is the sender. She can decide how thoroughly to investigate the borrower, but must communicate the resulting assessment through the bank's fixed rating system. In Figure~\ref{fig:demo_utility-3}, we depict the scenario where three grades allow her to target both approval regions while using a residual rejection report to satisfy Bayes plausibility. In Figure~\ref{fig:demo_utility-2}, there are only two grades available. So, one approval target must be pooled with residual uncertainty. The posterior support can no longer separately reach both favorable regions while averaging back to the prior. This is why coarser reporting lowers the expected volume of approved loans.

In Section~\ref{sec:threshold}, we apply this framework to interesting economic settings, develop a tractable belief-threshold benchmark and discuss other applications. The extension with heterogeneous thresholds and sender values, which generalizes the simplified commercial lending example above to a more realistic setting, is analyzed in Section~\ref{sec:threshold-extend}.

\medskip
\noindent
\textbf{Affine independence.} We proceed with describing the theoretical properties of optimal information structures which allow us to provide alternative characterizations of the sender's problem. A finite collection of posteriors is \emph{affinely independent} if it contains no redundancy, meaning that no posterior can be reconstructed as a weighted average of the others with weights summing to one.\footnote{Formally, $\{\mu_1,\ldots,\mu_\ell\}$ is affinely independent if $\{\mu_i-\mu_1\}_{i=2}^{\ell}$ are linearly independent.} The following result shows that we may restrict attention to information structures whose support is non-redundant in this sense.

\begin{Lemma}\label{Lemma:affdept}
There exists an optimal information structure whose support is affinely independent.
\end{Lemma}
If induced posteriors are affinely dependent, then Bayes plausibility does not uniquely pin down the message
probabilities assigned to those posteriors. More precisely, there is a family of probability
vectors $\tau$ over the given posteriors $\mu$ such that Bayes plausibility is satisfied. Moving probabilities within
this family changes the sender's expected payoff linearly, since posteriors and resulting actions are held fixed. Hence a weakly optimal
choice is attained at an extreme point of this family, where at least one posterior is assigned zero probability. Equivalently, one can
drop a posterior and redistribute probabilities of posteriors without lowering payoff.

\medskip
\noindent
\textbf{Compress--then-Concavify.} Affine independence of optimal information structures implies the sender only needs to vary beliefs in independent directions. With $k$ messages there could be at most $k\!-\!1$ independent directions away from the prior. Therefore, the sender can equivalently \emph{compress} the original state space to a \textit{pseudo--state space} of size at most $k$ and work with beliefs on this $k$-element simplex.

Formally, let \(\mathcal L_{k-1}(\mu_0)\) denote the collection of \((k{-}1)\)-dimensional affine subspaces \(L\) passing through \(\mu_0\).\footnote{This collection is canonically
identified with a \textit{Grassmannian}. Indeed, every such affine subspace can be written uniquely as \(L=\mu_0+ \widehat{L}\), where \(\widehat L\) is a \((k{-}1)\)-dimensional linear subspace of the translation space of $\mathbb{R}^{|\Omega|}$. Thus \(\mathcal L_{k-1}(\mu_0)\) is the Grassmannian of \((k{-}1)\)-planes through \(\mu_0\), so the notation is well defined. If $\mu_0$ is interior, then \(\Delta(\Omega)\cap L\) has relative dimension \(k{-}1\) for every \(L\in\mathcal L_{k-1}(\mu_0)\), otherwise the dimension may be smaller.} For a given \(L\in\mathcal L_{k-1}(\mu_0)\),  the slice \(\Delta(\Omega)\cap L\)
is a \((k{-}1)\)-dimensional polytope.

Define the associated \emph{pseudo-state space} by the extreme points of this slice $\Omega^*:=\Ext\bigl(\Delta(\Omega)\cap L\bigr).$ Since every point of a polytope is a convex combination of its extreme points, every posterior in the slice can be represented as a belief over the compressed state space \(\Omega^*\).  Therefore, by Carath\'eodory's theorem, any point in the concave envelope on this slice can be implemented with at most \(k\) posteriors. Hence restricting persuasion to \(\Delta(\Omega)\cap L\) is equivalent to studying a canonical persuasion problem on \(\Omega^*\). This yields a familiar representation of the sender's value, which we call \textit{compress--then-concavify}.

\begin{Proposition}\label{Optimal Summary} For any $k \geq 2$ and $\mu_0 \in \Delta(\Omega)$
\[
V(k,\mu_0)
=
\sup_{L\in\mathcal L_{k-1}(\mu_0)}
\Bigl(\operatorname{cav}\bigl(\hat u^S\big\vert_{\Delta(\Omega)\cap L}\bigr)\Bigr)(\mu_0).
\]
\end{Proposition}

 This viewpoint also clarifies the idea behind $k$-concavification, and provides a clear geometric intuition.  For $k=2$, once we select a one-dimensional affine subspace $L$ (a line passing through the prior $\mu_0$) and extend it to the boundary of the simplex, the intersection points (extreme points) identify the \textit{pseudo-state space} $\Omega^*$. The problem becomes a persuasion problem over a binary state space and it can be solved by concavifying the payoff function $\hat{u}^S$ over $\Delta(\Omega^*)$. In the illustration above, Figure~\ref{fig:demo_utility-2} shows exactly this reduction: the sender's payoff is concavified on the extended line segment containing the two-posterior optimum.

\medskip
\noindent
\textbf{Implications for the belief based approach.}
The optimal compression approach allows us to use our findings in other settings employing belief-based analysis. Generally, the constraints on the size of the message space can be transformed to constraints on the dimensionality of the subspace where the sender can provide information and induce posterior beliefs. In Appendix \ref{cheaptalk}, we apply this observation to a model of cheap talk with state-independent sender preferences by \cite{lipnowski2020cheap} where the sender's best equilibrium payoff is characterized by the \textit{quasi-concave envelope} of her interim value. With $|M|\le k$, feasible posteriors lie in some $L\in\mathcal L_{k-1}(\mu_0)$, so one can choose $L$ and then apply the quasi-concave envelope on $\Delta(\Omega)\cap L$.

\section{Value of Communication Capacity}
In this section, we quantify the sender’s \textit{value for additional communication capacity}. Formally, the value of the $k$\textsuperscript{th} message is the incremental gain from expanding capacity from $k - 1$ to $k$ messages,
\[
\Delta_k := \VS(k,\mu_0)-\VS(k-1,\mu_0).
\]
Two immediate implications of $k$-concavification are \textit{monotonicity} and \textit{eventual saturation}. First, enlarging the message constraint can only expand the feasible set, so $\Delta_{k}\ge 0$. If satisfying Bayes plausibility forces the sender to induce a posterior leading to a low-payoff action, then the sender can gain from an additional message that lets her reduce the probability mass on that action.  Second, the sender eventually attains the unconstrained optimum. If \(k \ge n\), the support constraint no longer binds, and therefore \(\Delta_{k+1}=0\) for all such \(k\).

\medskip
\noindent
\textbf{Value of an additional message.} We first establish a bound on how much the sender can gain from an extra message across all persuasion problems. In stating the bound we normalize the sender's interim payoff so that $\min_{\mu}\hat u^S(\mu)=0$. This is without loss of generality because adding a constant to $\hat u^S$ does not affect optimal information structures. The general statement is provided in the proof.

\begin{theorem}\label{theorem:lowerbound}
For every $k$ and prior $\mu_0$,
\[
\Delta_{k+1} \le \frac{2}{k+1} V\left(k+1, \mu_0\right) \leq \frac{2}{k-1} V\left(k, \mu_0\right).
\]
Moreover, the bounds are tight as stated in Corollary \ref{corollary:tight}.
\end{theorem}

The result reveals several interesting and general properties about coarse communication. The upper bound on the value of an additional message becomes tighter as the communication capacity grows. The $(k{+}1)$\textsuperscript{th} message can increase the sender's payoff by at most a $\frac{2}{k-1}$ fraction of what she already achieves with $k$ messages. The maximum percentage gain can be large for small $k$ and it decays at rate $O(1/k)$ as $k$ increases.

A useful interpretation is in terms of a discrete elasticity of the sender's payoff with respect to message capacity. Increasing capacity from $k$ to $k+1$ is a $1/k$ proportional increase in the number of available messages, resulting in an elasticity bound of
\(
\frac{\Delta_{k+1}/V(k,\mu_0)}{1/k} \le \frac{2k}{k-1}.
\)
This elasticity bound decreases in $k$ and converges to $2$ as $k\to\infty$.

The inequality expressed in terms of \(V(k+1,\mu_0)\) has an interpretation as the cost of losing the \((k+1)\)\textsuperscript{st} message. It says that removing access to this additional message reduces the sender's payoff by at most a \(\frac{2}{k+1}\) fraction of the payoff achievable when \((k+1)\) messages are available.

The constants $\frac{2}{k-1}$ and $\frac{2}{k+1}$ are sharp.  Moreover, when the bound is tight in the sense that either inequality binds, then in fact \emph{both} inequalities bind and all expressions hold with equality. We postpone the discussion to  Section~\ref{sec:threshold} where we identify environments that attain the bound, where Corollary~\ref{corollary:tight} provides an exact-equality class in a broader event-threshold environment.

The proof of Theorem \ref{theorem:lowerbound} involves constructing alternative $k$-dimensional feasible posterior distributions that are derived from the $(k+1)$-optimal posterior distribution, $\tau^*_{k+1}$. To do so, we generate $k+1$ distinct $k$-dimensional information structures by averaging pairs of posteriors from the support of $\tau^*_{k+1}$, while keeping the remaining posteriors unchanged.

Specifically, we take two arbitrary posteriors, $\mu_m$ and $\mu_{m'}$ within the support of $\tau^*_{k+1}$, and merge them into a single posterior, $\mu_{m,m'}$ by weighting them with their relative ex-ante probabilities.\footnote{This new posterior is calculated as follows: $\mu_{m,m'}:=\frac{\tau(\mu_m)}{\tau(\mu_m)+\tau(\mu_{m'})} \mu_m + \frac{\tau(\mu_{m'})}{\tau(\mu_m)+\tau(\mu_{m'})} \mu_{m'}$.}  Consequently, the constructed $k$-dimensional plausible posterior distribution, $\tau_k$, assigns the probability $\tau_k(\mu_{m,m'})=\tau^*_{k+1}(\mu_m) + \tau^*_{k+1}(\mu_{m'})$ and $\tau_k(\mu_{m''})=\tau^*_{k+1}(\mu_{m''})$ for every $m'' \in M \setminus \{m,m'\}$. The utilities provided by these new information structures are linked to $V(k+1,\mu_0)$, because they contain $k-1$ posteriors which are also in the support of $\tau^*_{k+1}$. Moreover, by the optimality of $\tau^*_{k}$, new posterior distributions $\tau_k$ must provide weakly less utility compared to $\tau^*_{k}$. Combining these observations, we establish the upper bound.

\medskip
\noindent
\textbf{Value of communication capacity.} Corollary~\ref{cor:scaling} is an \textit{integrated} version of Theorem~\ref{theorem:lowerbound}.
Rather than bounding the gain from a single extra message, it bounds how the entire
value function scales as we move from one capacity level to another.
\begin{corollary}\label{cor:scaling}
Fix a prior $\mu_0$. Let $\ell,k,m$ be integers with $2\le \ell \le k \le m$.
\[
\frac{k(k-1)}{\ell(\ell-1)} V(\ell,\mu_0)\ \ge\ V(k,\mu_0)\ \ge\ \frac{k(k-1)}{m(m-1)} V(m,\mu_0).
\]
Moreover,
\[
0\ \le\ V(k,\mu_0)-V(\ell,\mu_0)
\ \le\ \Bigl(1-\frac{\ell(\ell-1)}{k(k-1)}\Bigr)V(k,\mu_0)
\ \le\ \Bigl(\frac{k(k-1)}{\ell(\ell-1)}-1\Bigr)V(\ell,\mu_0).
\]
\end{corollary}

A convenient way to read the corollary is to define the normalized value $W(k):=\frac{V(k,\mu_0)}{k(k-1)}.$ Since $k(k-1)$ is the number of pairs of distinct messages, $W(k)$ can be interpreted as the sender's value per available pairwise distinction in a $k$-message language. The first inequality chain implies that $W(k)$ is weakly decreasing in $k$. Equivalently, moving from $\ell$ to $k$ messages scales the value by at most $\frac{k(k-1)}{\ell(\ell-1)}$, while $k$ messages guarantee at least the fraction $\frac{k(k-1)}{m(m-1)}$ of what is achievable with $m$ messages.

The second inequality chain bounds the incremental value of expanding capacity from $\ell$ to $k$
messages. In particular, it bounds the gain $V(k,\mu_0)-V(\ell,\mu_0)$ both as a fraction of the $k$-message payoff and as a multiple of the $\ell$-message payoff.

\medskip
\noindent
\textbf{Sufficient v. binary commmunication capacity.}  It is especially useful to compare the no-information benchmark
$V(1,\mu_0)=\hat u^S(\mu_0)$, binary capacity, and sufficient capacity. Setting $\ell=2$ and $m=n$
in Corollary~\ref{cor:scaling} yields a direct comparison between binary and sufficient communication, and it
also isolates the gap between the first nontrivial gain
$\Delta_2 = V(2,\mu_0)-V(1,\mu_0)$ and the total gain to sufficient capacity
$\Delta_n = V(n,\mu_0)-V(1,\mu_0)$. We state this benchmark next.

\begin{corollary}\label{corolla:bound}
For every $k\ge2$ and $\mu_0\in\Delta(\Omega)$,
\[
\frac{k(k-1)}{2}\VS(2,\mu_0) \ \geq \ \VS(k,\mu_0)  \ \ge\ \frac{k(k-1)}{n(n-1)}\VS(n,\mu_0).
\]
Moreover,
\[
 V(n,\mu_0)-V(2,\mu_0)
\le \frac{(n-2)(n+1)}{n(n-1)} V(n,\mu_0)
\le \frac{(n-2)(n+1)}{2} V(2,\mu_0).
\]
\end{corollary}

The upper bound shows that a small increase in the number of messages beyond a binary setup can substantially improve the sender's payoff. Conversely, the lower bound characterizes the fraction of the full-information payoff that can be secured under a \(k\)-message constraint. For a larger state space, a smaller share of the total potential payoff is captured by \(k\) messages.\footnote{One can always inflate \(n\) by adding payoff-irrelevant states. A state is payoff irrelevant if it can be removed and its probability reassigned across the remaining states in a fixed way so that, for every action the receiver might choose under some belief, both players’ expected payoffs are unchanged. Such a state does not affect the receiver’s or the sender’s behavior. Thus one can first reduce the state space by deleting payoff-irrelevant states and pushing the prior forward accordingly, which yields a tighter statement of the bound.}

The second part of the corollary bounds the residual value left unexploited under
binary communication, $V(n,\mu_0)-V(2,\mu_0)$.  The first inequality shows that the residual is at most $\frac{(n-2)(n+1)}{n(n-1)}$ fraction of the full persuasion payoff. The second inequality bounds the residual by $\frac{(n+1)(n-2)}{2}$ times the binary payoff.

\begin{figure}[h!]
\centering
\includegraphics[width=.85\textwidth]{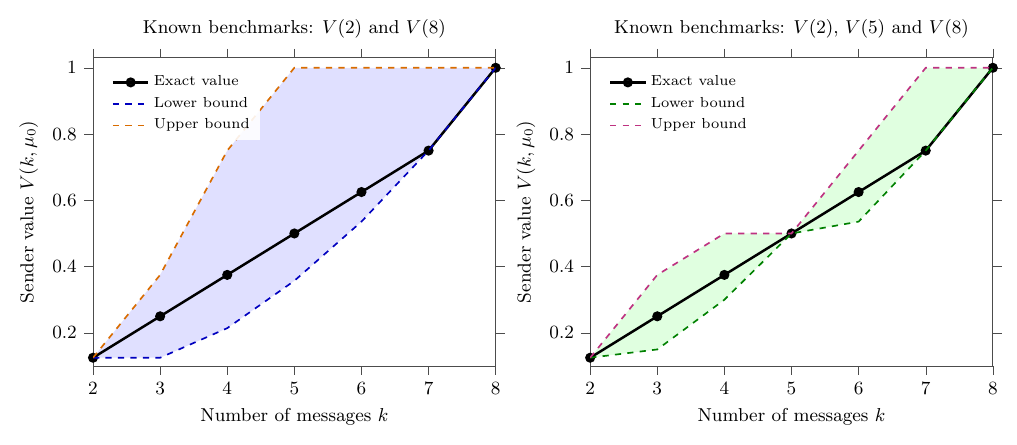}
\caption{The left panel uses only the values \(V(2,\mu_0)=1/8\) and \(V(8,\mu_0)=1\). The right panel adds  \(V(5,\mu_0)=1/2\). The exact value path is \(V(k,\mu_0)=(k-1)/8\) for $k<8$}
\label{fig:theorem-example}
\end{figure}

\medskip
\noindent
\textbf{Illustration of the bounds.} The general bounds are useful when endpoint capacities are tractable but intermediate capacities are not.
To illustrate, suppose that eight messages already suffice to attain the full sender payoff and binary communication value is known. Then Corollary~\ref{cor:scaling}, together with monotonicity, implies that for every \(k=2,\dots,8\),
\[
\max\!\left\{V(2,\mu_0),\frac{k(k-1)}{56}V(8,\mu_0)\right\}
\;\le\;
V(k,\mu_0)
\;\le\;
\min\!\left\{V(8,\mu_0),\frac{k(k-1)}{2}V(2,\mu_0)\right\}.
\]
Moreover, once the value at one intermediate capacity is also known, the same argument can be applied separately on the intervals to its left and right, yielding a strictly tighter admissible region for the remaining message levels.  Figure~\ref{fig:theorem-example} illustrates both cases, and also the tightness of the lower bound at $k=8$.\footnote{The exact values in the plot follow from the results discussed in Section~\ref{sec:threshold}, using an example from the same family employed in the tightness discussion.}

\section{Belief-Threshold Games}\label{sec:threshold}
While Theorem~\ref{theorem:lowerbound} is universal, additional primitives can yield stronger conclusions. In this section we focus on a special class of persuasion problems and characterize the value of an additional message explicitly.

Consider a setting in which the sender's payoff depends only  on the action, and not on the realized state. The receiver has a default action that is least preferred by the sender, and chooses to deviate from it only if sufficiently persuaded. This captures environments in which the sender prefers action over inaction, while the receiver requires enough confidence before abandoning a safe status quo. Versions of these preferences are widely utilized in the strategic communication literature to model environments where the receiver possesses a viable outside option such as lobbying \citep{lipnowski2020cheap} and seller-buyer interactions \citep{chakraborty2010persuasion}.

Formally, let $\Omega=\{\omega_1,\dots,\omega_n\}$ and $A=\{a_0,a_1,\dots,a_n\}$. The receiver's utility is defined using a single belief threshold $T\in(0,1)$ as follows:
\[
  u^R(a,\omega_j)=
  \begin{cases}
    0 & \text{if }  a=a_0\\[2pt]
    \frac{1}{T}-1 & \text{if }  a=a_i,\ j=i,\\[2pt]
    -1 &\text{if } a=a_i,\ j\neq i
  \end{cases} .
\]
Under this preference structure, the receiver chooses action \(a_i\) only if the posterior probability of state \(\omega_i\) is at least \(T\). Otherwise, the receiver selects the default action \(a_0\). Thus, a higher \(T\) means the receiver is more cautious about departing from the default, since it makes matching the correct state less rewarding. Define the set of posteriors where the receiver chooses a risky action as \(R_i := \{\mu \in \Delta(\Omega): \mu(\omega_i) \geq T\}\).

The sender's payoff depends only on the receiver’s action: \(u^S(a_0,\omega)=0\) and \(u^S(a_i,\omega)=1\) for all \(i \geq 1\). Hence, she aims to persuade the receiver away from the default action, making \(T\) a natural measure of the difficulty faced by the sender in influencing the receiver.  The sender’s payoff at any posterior \(\mu\) can be succinctly written as \(\hat{u}^S(\mu)=\mathbf{1}\{\mu \in \bigcup_{i=1}^n R_i\}\). With \(k\) messages, the sender’s maximum attainable payoff is the greatest ex-ante probability that the induced posteriors are in $R_i$, given Bayes plausibility. When there is no constraint on the number of messages, the sender can  provide full information, thereby allowing the receiver to perfectly match risky actions to states.

The baseline preferences should be read as a benchmark in which the sender mainly values movement away from the default, while the receiver primarily cares about matching the state with the correct action. Section~\ref{sec:threshold-extend} relaxes this benchmark by allowing heterogeneous thresholds and heterogeneous sender values across actions.\footnote{This formulation is more general than restricting each action to be optimal in a single state for the receiver. What matters is not that each risky action correspond to a single state, but that its comparison with the default be governed by a single event-threshold condition. Appendix~\ref{app:threshold-generality} shows that an action-versus-default comparison has this form if and only if the corresponding payoff difference is two-valued on some event and its complement in the state space.  Our baseline model is the special case in which each event is a singleton state and all thresholds are identical.}

\subsection{Optimal Information Structure}
Let $p_{(1)}\ge p_{(2)}\ge \cdots\ge p_{(n)}$ be the ordered entries
of $\mu_0$, and define their partial sum
\(
S_\ell := \sum_{i=1}^{\ell} p_{(i)}.
\)
We establish an upper bound on the sender's achievable payoff with \( k \) messages.

\begin{Lemma}\label{lem:UB-total-final}
For every $k$ and $\mu_0 \in \Delta(\Omega)$
\[
V(k,\mu_0)
\le
\begin{cases}
1, & S_k \ge T,\\[6pt]
\dfrac{S_{k-1}}{T}, & S_k < T.
\end{cases}
\]
\end{Lemma}

\noindent
\textbf{Accounting identity.} The argument is an accounting exercise. Each posterior that induces a risky action must place at least \(T\) probability on some state, while the prior can supply only a limited amount of mass across such states.

To see this, fix a feasible information structure $\tau=\{(\tau_i,\mu_i)\}_{i=1}^k \in \mathcal I(k,\mu_0).$ Let $\mathcal R := \bigcup_j R_j$ and $v(\tau):=\sum_{i:\,\mu_i\in\mathcal R}\tau_i,$ so \(v(\tau)\) is the ex-ante probability that the receiver takes a risky action under \(\tau\). Also let $r:=\bigl|\{i:\mu_i\in\mathcal R\}\bigr|$ be the number of induced posteriors that lead to a risky action.

For each \(i\) with \(\mu_i\in\mathcal R\), choose a state \(\omega_i\) such that $\mu_i(\omega_i)\ge T,$ and define $J:=\{\omega_i:\mu_i\in\mathcal R\}.$ Since distinct risky posteriors may correspond to the same state, we have \(|J|\le r\). Now we state the accounting identity:
\begin{align*}
T\,v(\tau) = T\sum_{i:\,\mu_i\in\mathcal R}\tau_i \le \sum_{i:\,\mu_i\in\mathcal R}\tau_i\,\mu_i(\omega_i) \le \sum_{\omega\in J}\sum_{m=1}^k \tau_m \mu_m(\omega) = \sum_{\omega\in J}\mu_0(\omega) \le S_{|J|}\le S_r .
\end{align*}
The first inequality uses \(\mu_i(\omega_i)\ge T\) for every risky posterior. The second enlarges each inner sum from the risky messages associated with \(\omega\) to all messages. Since all terms are nonnegative, this can only increase the expression. The following equality is Bayes plausibility. Finally, \(\sum_{\omega\in J}\mu_0(\omega)\le S_{|J|}\) because \(S_t\) is the sum of the \(t\) largest coordinates of the prior, and \(|J|\le r\).

Therefore every feasible information structure \(\tau\) satisfies $v(\tau)\le \frac{S_r}{T}.$ Since \(r\le k\), this implies $v(\tau)\le \frac{S_k}{T}.$ Taking the supremum over feasible information structures, \(V(k,\mu_0)=\sup_{\tau\in \mathcal I(k,\mu_0)} v(\tau)\), we obtain \(V(k,\mu_0) \le \frac{S_k}{T}\).  When \(S_k\ge T\), this upper bound is vacuous.
When \(S_k<T\), \(r=k\) is impossible. Hence \(r\le k-1\), and we obtain \(V(k,\mu_0)\le \frac{S_{k-1}}{T}\).

\medskip
\noindent
\textbf{Probability budget.} A useful way to read the accounting identity is as a \emph{probability-budget} interpretation. To induce any risky action, a message must allocate at least \(T\) posterior probability. Equivalently, each risky message must \emph{spend} at least \(T\) units of posterior mass on some state, while Bayes plausibility caps the total mass available across any \(r\) targeted states at \(S_r\). Hence when \(S_k<T\) there is not enough prior mass to fund \(k\) risky messages, and at least one must be safe.

\medskip
\noindent
\textbf{Construction of the optimal information structure.} The tractability of belief-threshold games allows us to give an explicit construction that achieves the upper bound established in Lemma~\ref{lem:UB-total-final}.

\begin{theorem} \label{threshold-prop}
For any $\mu_0$, $T$, and $k$,
there exists a Bayes plausible information structure that attains the bound identified in Lemma \ref{lem:UB-total-final}.
\end{theorem}

Theorem~\ref{threshold-prop} shows that the sender should use the \(k\) available messages to create persuasive posterior regions rather than to identify the state as finely as possible. The optimal design assigns persuasive messages to the states with the largest prior probabilities.  When \(S_k\ge T\), all \(k\) messages can be made persuasive by assigning them to the \(k\) most likely states and pooling the remaining states into those messages. When \(S_k<T\), only \(k-1\) messages can induce a risky action, so the sender targets the \(k-1\) most likely states and sets each persuasive posterior exactly at the threshold, \(\mu_i(\omega_{(i)})=T\). Any posterior above \(T\) is slack---it does not change the receiver's action and lowers the ex-ante frequency with which persuasive messages can be sent. The remaining states are pooled into a residual message that induces the safe action.

\medskip
\noindent
\textbf{Case 1: Full value is feasible.} When $S_k \geq T$, attaining the upper bound in Lemma \ref{lem:UB-total-final} requires that all $k$ messages induce a risky action.  The simplest Blackwell experiment that achieves this separates the most likely $k$ states from one another, and pools each most likely $k$ state with the least likely $n-k$ states.

More precisely, in state $\omega_{(i)}$ with $i \leq k$ the sender sends message $m_i$ with probability one, and never sends $m_i$ in any other most likely $k$ states. For tail states $\omega_{(\ell)}$ with $\ell>k$, she sends $m_i$ with probability $\frac{p_{(i)}}{S_k}$.

Bayes' rule then gives posterior weight
$\tau_i = p_{(i)}+\sum_{\ell>k} p_{(\ell)} \frac{p_{(i)}}{S_k}=\frac{p_{(i)}}{S_k}$, and posterior $\mu_i(\omega_{(i)}) = S_k$, zero on the other most likely $k$ states and the tail preserved in prior proportions. Because $S_k \geq T$, each message induces a risky action and the ex-ante cumulative probability of a risky action is one.

We summarize the resulting information structure below.

\begin{enumerate}
\item \textbf{Posteriors.} For each $i=1,\dots,k$:
\[
\mu_i (\omega_{(\ell)}) = \begin{cases}
    S_k  & \text{if } \ell =i \\
    0   &\text{if } \ell \leq k \text{ and }  \ell \neq i \\
    p_{(\ell)} & \text{if } \ell > k
\end{cases}
\]
\item \textbf{Posterior Weights.} $\displaystyle \tau_i:=\frac{p_{(i)}}{S_k}$ for $i=1,\dots,k$.
\end{enumerate}

\medskip
\noindent
\textbf{Case 2: At least one safe message is necessary.} When $S_k<T$, the bound in Lemma \ref{lem:UB-total-final} implies that at most $k-1$ messages may induce risky actions and the maximum total probability of those $k-1$ messages is $\frac{S_{k-1}}{T}$.

Constructing a Blackwell experiment that attains this bound follows a similar logic. It separates each of the most likely $k-1$ states from the others while pooling each with the $k-1$ tail. However, since the mass on most likely states is smaller, the optimal information structure now pools the most likely $k-1$ states with the tail just enough to attain the belief threshold. The final $k$\textsuperscript{th} message induces a safe action to attain Bayes plausibility.

Formally, in state $\omega_{(i)}$ for $i \leq k-1$ the sender sends $m_i$ with probability one. In any tail state $\omega_{(\ell)}$ ($\ell \geq k$) she sends $m_i$ with probability $\frac{p_{(i)}(1-T)}{T(1-S_{k-1})}$ for each $i \leq k-1$, and she sends the residual message $m_k$ with probability $\frac{T-S_{k-1}}{T(1-S_{k-1})}$.

Bayes' rule  yields $\tau_i=\frac{p_{(i)}}{T}$, and conditional on $m_i$, the posterior has $\mu_i\left(\omega_{(i)} \right)=T$, zero on the other top coordinates, and the tail in prior proportions.

For the message $m_k$, we have $\tau_k=\frac{T-S_{k-1}}{T}$ and the posterior places all probability on the tail in prior proportions and induces a safe action.

Only the first $k-1$ messages are risky, and the total ex-ante probability of risky messages is $\sum_{i=1}^{k-1} \tau_i =\frac{S_{k-1}}{T}$. The resulting information structure attains the upper bound identified in Lemma \ref{lem:UB-total-final}. We summarize the resulting information structure below.

\begin{figure}[htbp]
\centering

\begin{subfigure}[t]{0.48\linewidth}
\centering
\includegraphics[scale=0.8]{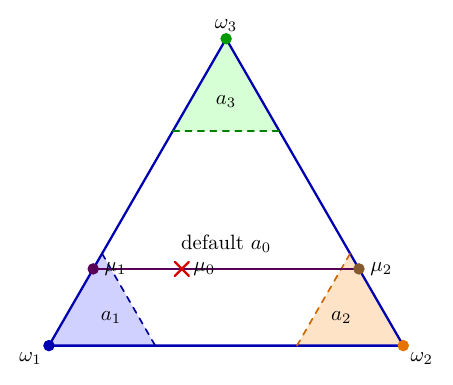}
\caption{Both messages persuasive}
\end{subfigure}
\hfill
\begin{subfigure}[t]{0.48\linewidth}
\includegraphics[scale=0.8]{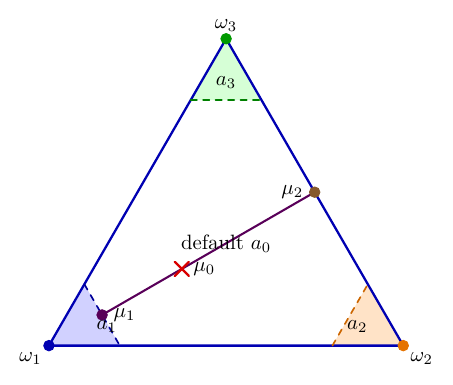}
\caption{One persuasive message}
\end{subfigure}

\caption{$\mu_0=(1/2,1/4,1/4)$, $k=2$ messages. Panel (a), $T=0.7$, and Panel (b) $T=0.8$}
\label{fig:three-state-threshold}
\end{figure}

\begin{enumerate}
\item\textbf{Posteriors.} For each $i=1,\dots,k-1$:
\[
\mu_i (\omega_{(\ell)}) = \begin{cases}
    T  & \text{if } \ell =i \\
    0   &\text{if } \ell \leq k-1 \text{ and }  \ell \neq i \\
    \frac{(1-T) p_{(\ell)} }{1-S_{k-1}}  & \text{if } \ell > k -1
\end{cases} \ \text{ and } \ \mu_k(\omega_{(\ell)}) = \begin{cases}
      0   &\text{if } \ell \leq k-1  \\
      \frac{p_{(\ell)}}{1-S_{k-1}} & \text{if } \ell > k-1
\end{cases}
\]

\item \textbf{Posterior Weights.} $ \tau_i:=\frac{p_{(i)}}{T}$ for $i \leq k-1$,
$ \tau_k:=1-\sum_{i=1}^{k-1}\tau_i=\frac{T-S_{k-1}}{T}>0$.
\end{enumerate}
Figure~\ref{fig:three-state-threshold} illustrates the geometry behind Theorem~\ref{threshold-prop} and the constructions above.\footnote{In panel(a), the optimal information structure induces posteriors $\mu_1=(0.75,0,0.25)$ and $\mu_2=(0,0.75,0.25)$ with weights $\tau_1=2/3$ and $\tau_2=1/3$, so both messages are persuasive. In panel (b), the optimal information structure induces $\mu_1=(0.8,0.1,0.1)$ and $\mu_2=(0,0.5,0.5)$ with weights $\tau_1=5/8$ and $\tau_2=3/8$.}

\medskip
\noindent
\textbf{Optimal compression.}
Theorem~\ref{threshold-prop} identifies the relevant optimally compressed pseudo-state space for the sender’s problem. The sender separates only the highest-prior states that can support persuasive posteriors and pools all lower-prior states into a tail that preserves their relative proportions. When \(S_k \ge T\), the capacity constraint is slack, because the top \(k\) states carry enough prior mass to absorb the tail without driving any targeted posterior below \(T\). The optimal compression therefore separates these \(k\) states and distributes the pooled tail across their messages so that all \(k\) posteriors remain persuasive, yielding sender value \(1\). When \(S_k < T\), by contrast, the capacity constraint binds. The sender cannot sustain \(k\) persuasive messages, so the optimal compression separates only the \(k-1\) highest-prior states, uses the pooled tail to construct \(k-1\) posteriors exactly at threshold \(T\), and assigns the remaining probability to a residual message that preserves the tail in prior proportions and induces the safe action. This compressed representation underlies the geometric analysis in Figure~\ref{fig:three-state-threshold}.

\medskip
\noindent
\textbf{Discussion and Examples} Theorem~\ref{threshold-prop} yields two key practical insights for constrained senders. First, scarce messages should be dedicated to the most probable favorable states (e.g., core customer segments or highly plausible policy scenarios), while unlikely tail contingencies are pooled into a residual category. Second, optimal persuasion strictly avoids over-persuasion. Posteriors are pushed exactly to the receiver's threshold $T$ and no further. This systematically under-communicates tail risks, highlighting a tension where more disclosure might be socially optimal but contradicts the sender's private incentives.

E-commerce platforms frequently face interface constraints, such as a single \textit{Buy Box} or a finite number of \textit{Top Pick} badges.\footnote{\citet{peitz2025biased} show that a monopolist using personalized recommendations under uniform pricing may optimally bias recommendations when consumers differ in how well informed they are about product values.} As our model predicts, there is evidence that platforms optimize these coarse signals by pooling \textit{perfect} product matches with highly profitable, \textit{just-good-enough} items. Perfectly identifying the best items leaves a slack posterior belief in terms of consumer confidence remaining unexploited. \cite{chen2021steering} empirically document this exact behavior: Amazon's Buy Box systematically steers consumers away from objective best offers (e.g., the lowest price for the same item) toward offers utilizing Amazon's own fulfillment services. By pooling these more profitable offers into the coarse Buy Box signal, the platform extracts maximum surplus while keeping the consumer's expected utility just high enough to induce a purchase.

Financial regulation often maps fine-grained information into a small set of rating categories, so small differences in underlying risk can have discontinuous effects on investment eligibility or collateral treatment \citep{barbosa2021public, bindseil2017eurosystem}. Ratings are often conditions for issuance, and underwriters can often anticipate rating-model outputs and revise deal structure iteratively to secure them. \citet{griffin2012did} show this in the \textit{collateralized debt obligation} (CDO) market. Using internal data from one major credit rating agency on 916 CDOs issued between 1997 and 2007, they find that the actual share of each deal rated AAA averaged 75.5\%, whereas the agency's internal model implied only 63.4\%.
They also show that credit rating agencies seem to disproportionately make ex-post adjustments that move CDOs from model-implied lower ratings into the higher AAA rating category.
For CDOs issued before April 1, 2007, only 1.3\% of AAA tranches satisfied the agency's published AAA default criterion, while 92.5\% satisfied only the published AA criterion. These findings are consistent with a sender operating in a coarse, threshold-based language and using discretion to expand the favorable category up to the indifference point of the decision rule, rather than leaving slack in the top category.

These patterns recur across lending, retail platforms, and financial regulation, or wherever a coarse menu of categories sits between rich underlying information and a threshold-based decision.

\subsection{Value of an Additional Message}
We next characterize how the sender’s payoff evolves with the number of available messages. Recall the notation
\(
\Delta_k := V(k, \mu_0) - V(k-1, \mu_0)
\).  Let
\(
k^\ast := \min\{k : S_k \ge T\}
\)
denote the first index at which it becomes feasible to construct an information structure with full value. Using Theorem \ref{threshold-prop} we can characterize the value of an additional message.

\begin{Proposition} \label{thm:mono}
The sequence $\left\{\Delta_k\right\}_{k=2}^{k^*-1}$ is weakly decreasing in $k$. Moreover, the cutoff increment $\Delta_{k^*}$ is smaller than $\Delta_{k^*-1}$ if and only if $T < S_{k^*-2} + p_{(k^*-2)}$ .
\end{Proposition}

An extra message does not simply generate \textit{more information} in the abstract. It gives the sender one more category with which to allocate scarce persuasive capacity across contingencies that must each clear an adoption standard. Before the cutoff \(k^\ast\), each new message is used to activate one additional risky action by targeting another state, so the gain from communication rises one message at a time. But these gains decline because the sender optimally targets the most likely states first, leaving progressively less likely states for later messages. Formally, before the cutoff $k^* \geq k$, the \(k\)th message lets the sender target one additional state—the next most likely one—and convert \(\frac{p_{(k-1)}}{T}\) more ex-ante probability from safe action to risky action. Thus, there are diminishing returns from expanding a coarse message space when the default is still unavoidable.

At the cutoff, the extra message has a \textit{last-mile} gain. It eliminates the default altogether by allowing all relevant posterior mass to be distributed across persuasive messages. So the gain \(\Delta_{k^\ast}=1-\frac{S_{k^\ast-2}}{T}\) is the leftover safe probability that was previously unavoidable. Whether this \textit{last-mile} gain is larger or smaller than the previous increment depends on how much residual safe mass was left right before the cutoff relative to the incremental mass from targeting the next state, yielding the condition \(T \lessgtr S_{k^\ast-2}+p_{(k^\ast-2)}\). Beyond the cutoff point, additional messages have no value because the sender already induces action with probability one.

\medskip
\noindent
\textbf{Tightness of Theorem \ref{theorem:lowerbound}.}
Corollary~\ref{corollary:tight} provides a sharpness result. The proof shows that equality in Theorem~\ref{theorem:lowerbound} is attained in a benchmark configuration with two features. First, the \((k+1)\)\textsuperscript{st} message is exactly pivotal, so allowing \(k+1\) rather than \(k\) messages eliminates the remaining default mass. Second, the threshold-relevant mass is spread evenly across the top \(k+1\) states. The first feature makes the $k+1$\textsuperscript{th} message especially influential, while the second feature keeps the \(k\)-message benchmark as low as possible and thereby maximizes the marginal gain. The construction intuitively maximizes the possible value of the $(k+1)$\textsuperscript{st} message.

\begin{corollary}
\label{corollary:tight}
The constants in Theorem~\ref{theorem:lowerbound} are sharp. In the baseline
belief-threshold model, for every \(k\le n-2\) there exists a prior \(\mu_0\) such that
\[
\Delta_{k+1}
=
\frac{2}{k+1}V(k+1,\mu_0)
=
\frac{2}{k-1}V(k,\mu_0).
\]
Hence the constants in Theorem~\ref{theorem:lowerbound} cannot be improved.
\end{corollary}

\medskip
\noindent
\textbf{Communication costs.}
No message system is costless. Even a minimal vocabulary requires attention, training, and organizational processing. Suppose there is a cost associated with a \(k\)-message system, which is weakly increasing in $k$ and discretely convex. Proposition~\ref{thm:mono} implies that \(V(k,\mu_0)\) is weakly concave (in discrete differences) on \(\{1,\dots,k^\ast-1\}\) and at \(k^\ast\) it may break the concavity. The problem therefore separates cleanly. One first solves the interior problem on \(\{1,\dots,k^\ast-1\}\) by comparing marginal values to marginal costs, and then compares that interior solution to \(k^\ast\).

This yields two practical implications. First, communication systems should be refined selectively rather than uniformly. As long as $k<k^\ast$, the marginal value of an additional message declines with $k$, so new labels, ratings, or recommendations should be allocated to the most common or most decision-relevant contingencies, while unlikely tail states remain pooled in a residual category. This sharpens a familiar trade-off in organizational economics: richer codes improve adaptation, but they are costly to create, teach, monitor, and process \citep{cremer2007language, sobel2015broad}. It also resonates with the bounded-rationality: coarse categories economize on cognitive burden at the cost of losing finer distinctions \citep{rubinstein2000economics, wilson2014bounded}.

Second, investments in communication are naturally lumpy. Away from the full-persuasion cutoff $k^\ast$, an extra message only marginally improves persuasion and should be adopted only when that gain exceeds the cost of a richer language. Near $k^\ast$, however, one final category can have a first-order effect by eliminating the default action altogether. This predicts that firms will often tolerate coarse pooling and expand their message space only when an additional category changes the downstream action rather than refining an already actionable classification.

\subsection{Comparative Statics}

\textbf{Concentration of the prior.} A more concentrated prior places relatively more probability on the most likely states, and therefore increases the partial sums of the ordered coordinates. We measure this notion of concentration by majorization.  We say that \(\mu_0\) majorizes \(\nu_0\) if $S_\ell(\mu_0)\ge S_\ell(\nu_0)$ for all $\ell=1,\dots,n.$

Greater prior concentration, in this sense, makes persuasion easier because it places more probability mass on the most likely contingencies---exactly the contingencies the sender wants to target first when communication is coarse. For a seller, this means that when consumers are already relatively likely to find a few products especially suitable, a small number of recommendation categories goes further. For a think tank, when a few policy contingencies are especially plausible, scarce messages can be used more effectively.

\begin{Proposition}\label{prop:majorization-new}
Let \(\mu_0,\nu_0\in\Delta(\Omega)\) and suppose that \(\mu_0\) majorizes \(\nu_0\). Then:
\begin{enumerate}
\item \(V(k,\mu_0) \ge V(k,\nu_0)\).
Equivalently, \(\mu\mapsto V(k,\mu)\) is symmetric and Schur-convex.

\item \(k^\ast(\mu_0) \le k^\ast(\nu_0)\).

\item If \(k^\ast(\mu_0)=k^\ast(\nu_0)=k^\ast\), then
\(
\Delta_{k^\ast}(\mu_0)
\le \Delta_{k^\ast}(\nu_0).
\)

\item If, in addition, \(p_{(k)}(\mu_0)\ge p_{(k)}(\nu_0)\)  for every
\(k<k^*(\mu_0),\) then \(
\Delta_k(\mu_0)=\tfrac{p_{(k-1)}(\mu_0)}{T}\ \ge\ \tfrac{p_{(k-1)}(\nu_0)}{T}=\Delta_k(\nu_0).
\)
\end{enumerate}
\end{Proposition}

\medskip
\noindent
\textbf{Threshold.}  A higher threshold \(T\) corresponds to a stricter standard for action---the buyer needs stronger confidence to purchase, or the policymaker needs stronger evidence to depart from the status quo. This lowers the sender's value for every fixed \(k\) and increases the number of messages needed to eliminate the default.  The effect on the marginal value of an extra message is non-monotone. An additional category is useless when existing capacity already suffices, most valuable when it is just enough to move the receiver from partial to full activation, and less valuable again when the standard is so demanding that even the extra category cannot do much. The returns to finer communication are largest near the receiver's decision margin, where one more distinction can convert persuasive potential into action.

\begin{Proposition} \label{prop-thresh}
    $V(k,\mu_0)$ is weakly decreasing in $T$. Moreover, $\Delta_{k}$ is increasing in $T$ if $T \in (S_{k-1}, S_k)$ and strictly decreasing if $T \geq S_k$.
\end{Proposition}

\subsection{Extensions}\label{sec:threshold-extend}

The baseline model extends in three natural directions: the receiver's threshold may vary by risky action, the sender may value only some risky actions, and sender values may differ across risky actions.

A single probability-budget notation summarizes the first two extensions and organizes the third. For each action \(a_i\), let \(c_i\) denote the maximum ex-ante probability with which action \(a_i\) can be induced. We refer to $c_i$ as the action's \textit{capacity}, refering to the probability budget. Let
\(c=(c_1,\ldots,c_m)\). We often use the decreasing rearrangement \(c_{(1)}\ge\cdots\ge c_{(m)}\). The  relevant partial sums are $C_\ell:=\sum_{r=1}^{\ell}c_{(r)}$ with the convention $C_0:=0,$ and \(C_\ell=C_m\) for \(\ell>m\).

Given the capacity vector \(c\), the equal-value threshold problem has value
\begin{equation}\label{eq:capacity-value}
\mathcal V_c(k):=
\begin{cases}
1, & C_k\ge 1,\\
C_{k-1}, & C_k<1.
\end{cases}
\end{equation}
For \(k_c^*:=\min\{\ell:C_\ell\ge1\}\), the same object gives the cutoff and marginal-value
statements. For \(k<k_c^*\), the \(k\)-th message contributes \(c_{(k-1)}\). At
\(k=k_c^*\), it contributes the remaining gap \(1-C_{k_c^*-2}\), and after the cutoff it has
no value.
The details and proofs are in Appendix~\ref{appendix:ext}.

\medskip
\noindent
\textbf{Heterogeneous belief thresholds.}
Let the receiver's threshold for action \(a_i\) be \(T_i\in(1/2,1)\), and suppose the sender
values all risky actions equally. In this problem the capacity becomes $c_i:=\frac{p_i}{T_i}$. A message inducing \(a_i\) must put at
least \(T_i\) posterior probability on \(\omega_i\), so action \(a_i\) can be induced with
ex-ante probability at most \(p_i/T_i\). Thus, the heterogeneous-threshold value follows from equation \eqref{eq:capacity-value} as $\mathcal V_c(k)$ with $c_i:=\frac{p_i}{T_i}$.

The priors and thresholds enter only through the ordered capacity profile \(c\). All baseline
comparative statics carry over by replacing the ordered prior sums with the ordered capacity
sums \(C_\ell\). The details and the construction of optimal information structure are in Appendix~\ref{sub-het-belief}.

\medskip
\noindent
\textbf{A valued subset of risky actions.}
Return to a common threshold \(T\), but suppose the sender benefits only from actions in
\(G\subseteq\{1,\ldots,n\}\). In this case, we track only the capacities of valuable actions and apply equation \eqref{eq:capacity-value}. So the value is $\mathcal V_{c^G}(k)$ where $c^G=\left(\frac{p_i}{T}\right)_{i\in G}$.

Only valuable states enter the probability budget. Hence enlarging \(G\) weakly raises value,
weakly lowers the number of messages needed to eliminate the default whenever full value is
attainable, and otherwise affects value only through the ordered valuable capacities. The
details and the construction of the optimal information structure are provided in Appendix~\ref{sub-valuedd}.

\medskip
\noindent
\textbf{Heterogeneous sender values.}
With arbitrary thresholds and sender values \(v_i\ge0\), we keep the capacity vector from heterogeneous belief thresholds: \(c_i=p_i/T_i\).  Since the sender now cares which risky actions are induced, \(\mathcal V_c\) no longer gives the sender's payoff by itself. It does, however, identify the same feasibility regimes.

If \(\mathcal V_c(k)<1\), the default cannot be eliminated with
\(k\) messages. The problem is separable. Defining
\(z_i:=v_i c_i\) and ordering \(z_{(1)}\ge z_{(2)}\ge\cdots\), the sender obtains $V(k,\mu_0)=\sum_{r=1}^{k-1}z_{(r)}.$

If \(\mathcal V_c(k)=1\), the selected risky actions must absorb one unit of probability, so the
unit-mass constraint couples the actions, and support selection becomes a knapsack-type selection problem with a cardinality constraint:
\[
\max_{x_i\ge0}\sum_i v_i x_i
\quad\text{subject to}\quad
x_i\le c_i\ \forall i,\qquad
\sum_i x_i=1,\qquad
|\{i:x_i>0\}|\le k.
\]

However, a closed form can be obtained when values and capacities are aligned in the sense that
\(v_1\ge\cdots\ge v_n\) and \(c_1\ge\cdots\ge c_n\).  Let
\(k_c^*:=\min\{\ell:C_\ell\ge1\}\). Then $V(k,\mu_0)=\sum_{i=1}^{k-1} z_i$ whenever $k<k_c^*$, while $V(k,\mu_0)=\sum_{i=1}^{k_c^*-1} z_i
+v_{k_c^*}\left(1-C_{k_c^*-1}\right)$ for \(k\ge k_c^*\). Thus, \(\mathcal V_c\) continues to locate the cutoff, while value-adjusted capacities determine which feasible risky actions are most attractive.  Appendix~\ref{sub:het-val} gives the detailed analysis.

\section{Conclusion}
When the sender communicates through a message space with cardinality \(k\), Bayes plausibility must be satisfied by an information structure with support of size at most \(k\). Our main general result is that the sender's value is characterized by \(k\)-concavification. We also show that there exists an optimal information structure with affinely independent support. This yields a compress--then--concavify representation: the sender chooses a \((k-1)\)-dimensional affine subspace through the prior, or equivalently a pseudo-state space of size at most \(k\), and then solves a standard persuasion problem on the resulting slice of the simplex. In the appendix, we apply the same logic to cheap talk with state-independent sender preferences.

We next quantify the value of communication capacity. Across all persuasion problems, the gain from a \((k+1)\)\textsuperscript{st} message satisfies a tight bound. The value of adding a $(k+1)$\textsuperscript{st} message is at most $\frac{2}{k-1}$ times the payoff already attainable with $k$ messages. This establishes a fundamental property of coarse communication: the maximum percentage returns to expanding communication capacity necessarily decay at a rate of $O(1/k)$. Integrating this bound yields scaling inequalities that compare any two capacity levels \(\ell\le k\le m\). Equivalently, \(V(k,\mu_0)/(k(k-1))\) is weakly decreasing in \(k\). As a special case, we compare binary communication to sufficient communication capacity and bound the residual value left unexploited under binary communication.

Finally, we study a realistic and practical class of games called belief-threshold games. In this class, persuasion is governed by a probability budget. When capacity binds, the sender cannot afford to over-persuade. Instead, optimal information structures have to ration probability mass. They target the most likely states and push posteriors exactly to the boundary of the receiver's action regions to eliminate slack. The sender strategically uses the surplus probability of highly likely states to pool less likely states, ensuring the combined coarse message just clears the receiver's threshold. A simple cutoff characterizes when extra capacity ceases to matter. Before this cutoff, the incremental value of a new message declines because it targets progressively less likely states. This budgeting logic also yields clean comparative statics: greater prior concentration expands the sender's effective persuasive budget, while higher receiver thresholds strictly shrink it.

We then extend the threshold analysis in three directions. With heterogeneous belief thresholds, priors and thresholds matter through the ordered action-specific persuasion capacities. With a valued subset of risky actions, the value depends only on the ordered prior masses of the valuable states. With heterogeneous sender values, the binding region is solved by ranking value-adjusted capacities, and under aligned values and capacities we obtain a closed-form characterization beyond the binding region as well.

The probability-budget idea suggests a broader class of tractable environments than the general case considered here. In the binding regime \(C_k<1\), the sender's problem reduces to choosing a support subject to per-action capacities, so the same logic should extend beyond constant payoffs to richer separable sender objectives, support restrictions such as group quotas or category caps, and endogenous communication costs.

Overall, the paper isolates how an exogenous constraint on message cardinality changes the geometry of persuasion, the value of communication capacity, and the form of optimal information structures.

\section*{Statements and Declarations}

\noindent\textbf{Funding.} The authors did not receive support from any organization for the submitted work.

\medskip
\noindent\textbf{Competing interests.} The authors have no relevant financial or non-financial interests to disclose.

\medskip
\noindent\textbf{Data availability.} Data sharing is not applicable to this article as no datasets were generated or analyzed during the current study.

\bibliographystyle{apsr}
\bibliography{references}

\appendix

\section{Proofs} \label{appendix:proofs}

\subsection{k-Concavification}
\begin{proof}[\textbf{Proof of Proposition \ref{theorem:Equivalence}.}]
By definition, \(\tau \in \mathcal I(k,\mu_0)\) is optimal if and only if \(\E_\tau[\hat u^S(\mu)] = V(k,\mu_0).\)
So it is enough to show that $V(k,\mu_0)=\sup\{z:(\mu_0,z)\in \cavk(\hat u^S)\}.$

Take any \(\tau \in \mathcal I(k,\mu_0)\). By definition, there exist \((\tau_i,\mu_i)_{i=1}^k\) such that
\(\tau_i\ge 0,\) \(\sum_{i=1}^k \tau_i=1,\)  \(\sum_{i=1}^k \tau_i\mu_i=\mu_0, \)
and \(\E_\tau[\hat u^S(\mu)] = \sum_{i=1}^k \tau_i \hat u^S(\mu_i).\) Hence, by the definition of \(\cavk(\hat u^S)\), $(\mu_0,\E_\tau[\hat u^S(\mu)])\in \cavk(\hat u^S).$ Therefore $V(k,\mu_0)\le \sup\{z:(\mu_0,z)\in \cavk(\hat u^S)\}.$

Conversely, take any \(z\) such that \((\mu_0,z)\in \cavk(\hat u^S)\). By definition of \(\cavk(\hat u^S)\), there exist \((\tau_i,\mu_i)_{i=1}^k\) such that \(\tau_i\ge 0,\) \(\sum_{i=1}^k \tau_i=1,\)  \(\sum_{i=1}^k \tau_i\mu_i=\mu_0, \) and $z\le \sum_{i=1}^k \tau_i \hat u^S(\mu_i).$

Let \(\tau \in \mathcal I(k,\mu_0)\) be the information structure that assigns probability \(\tau_i\) to \(\mu_i\), for \(i=1,\dots,k\). Then \(\E_\tau[\hat u^S(\mu)] = \sum_{i=1}^k \tau_i \hat u^S(\mu_i)\ge z.\) Since this holds for every \(z\) with \((\mu_0,z)\in \cavk(\hat u^S)\), we obtain $V(k,\mu_0)\ge \sup\{z:(\mu_0,z)\in \cavk(\hat u^S)\}.$
Combining the two inequalities gives
\[
V(k,\mu_0)=\sup\{z:(\mu_0,z)\in \cavk(\hat u^S)\}.
\]
\end{proof}

\subsection{Optimal Compression}\label{compress}

\begin{proof}[\textbf{Proof of Lemma \ref{Lemma:affdept}}]
Among all optimal information structures, choose $\tau$ with minimal support, and let $\supp(\tau)=\{\mu_1,\dots,\mu_\ell\}$.

Suppose $\supp(\tau)$ is affinely dependent. Then there exists $(\lambda_1,\dots,\lambda_\ell)\neq 0$ such that $\sum_{i=1}^{\ell}\lambda_i=0$ and $\sum_{i=1}^{\ell}\lambda_i\mu_i=0.$

For each $t\in\mathbb{R}$, define $\tau_t(\mu_i):=\tau(\mu_i)+t\lambda_i$ for $i=1,\dots,\ell.$ Since $\sum_{i=1}^{\ell}\lambda_i=0$, we have $\sum_{i=1}^{\ell}\tau_t(\mu_i)=1.$ Since $\tau$ is Bayes plausible and $\sum_{i=1}^{\ell}\lambda_i\mu_i=0$,
\[
\sum_{i=1}^{\ell}\tau_t(\mu_i)\mu_i
=
\sum_{i=1}^{\ell}\tau(\mu_i)\mu_i
+
t\sum_{i=1}^{\ell}\lambda_i\mu_i
=
\mu_0.
\]
Hence every $\tau_t$ with nonnegative weights is Bayes plausible.

Let $\underline t:=\max_{\lambda_i>0}\left(-\frac{\tau(\mu_i)}{\lambda_i}\right),$ and  $\bar t:=\min_{\lambda_i<0}\left(-\frac{\tau(\mu_i)}{\lambda_i}\right).$ Because $(\lambda_1,\dots,\lambda_\ell)\neq 0$ and $\sum_{i=1}^{\ell}\lambda_i=0$, some $\lambda_i$ are positive and some are negative. Therefore, $\underline t<0<\bar t,$ and $\tau_t(\mu_i)\geq 0$ for all $i$ whenever $t\in[\underline t,\bar t]$. At $t=\underline t$ or $t=\bar t$, at least one posterior receives probability zero, so the support is strictly smaller.

Finally,
\[
\E_{\tau_t}\hat u^S
=
\sum_{i=1}^{\ell}\tau_t(\mu_i)\hat u^S(\mu_i)
=
\sum_{i=1}^{\ell}\tau(\mu_i)\hat u^S(\mu_i)
+
t\sum_{i=1}^{\ell}\lambda_i\hat u^S(\mu_i),
\]
so $\E_{\tau_t}\hat u^S$ is affine in $t$.

Since $\tau=\tau_0$ is optimal, $\E_{\tau_0}\hat u^S$ is maximal on $[\underline t,\bar t]$. An affine function on a compact interval attains its maximum at an endpoint, so at least one of $\tau_{\underline t}$ or $\tau_{\bar t}$ is also optimal. But each has strictly smaller support than $\tau$, contradicting the minimality of $|\supp(\tau)|$.

Therefore, $\supp(\tau)$ is affinely independent.
\end{proof}

Any information structure in \(\mathcal I(k,\mu_0)\) is supported on an affine subspace
through \(\mu_0\) of dimension at most \(k-1\). This suggests viewing the sender's problem
as first choosing a \((k-1)\)-dimensional slice of the belief simplex and then solving
persuasion on that slice. Formally, define
\[
\mathcal L_{k-1}
:=
\bigl\{
L\subseteq \operatorname{aff}(\Delta(\Omega))
\;\bigm|\;
\mu_0\in L,\ \dim L = k-1
\bigr\}.
\]
For each \(L\in\mathcal L_{k-1}\), let
\[
K_L:=\Delta(\Omega)\cap L
\]
and
\[
\mathcal I_k(L,\mu_0)
:=
\Bigl\{
\tau \in \Delta(\Delta(\Omega))
\;\Bigm|\;
\E_\tau[\mu]=\mu_0,\
\supp(\tau)\subseteq K_L,\
|\supp(\tau)|\le k
\Bigr\}.
\]

The set \(K_L\) is a compact convex polytope of affine dimension at most \(k-1\).
Its extreme points can be interpreted as pseudo-states. Although \(K_L\) may have
more than \(k\) extreme points, what matters is its affine dimension: standard
support-reduction implies that an optimal Bayes plausible distribution on \(K_L\)
can be chosen with at most \(k\) support points.

\begin{proof}[\textbf{Proof of Proposition \ref{Optimal Summary}.}]

\ \ \
\medskip
\noindent
\textbf{Step 1.} Define  $W(L):=
\sup_{\tau\in \mathcal I_k(L,\mu_0)} \E_\tau[\hat u^S(\mu)]$. We first show that $V(k,\mu_0)
=
\sup_{L\in\mathcal L_{k-1}(\mu_0)} W(L)$

Take any \(\tau\in \mathcal I(k,\mu_0)\). Since \(\E_\tau[\mu]=\mu_0\), we have $\mu_0\in \operatorname{conv}(\operatorname{supp}\tau)\subseteq \operatorname{aff}(\operatorname{supp}\tau).$ Hence \(\operatorname{aff}(\operatorname{supp}\tau)\) is an affine subspace of
\(\operatorname{aff}(\Delta(\Omega))\) containing \(\mu_0\). Moreover, $\dim \operatorname{aff}(\operatorname{supp}\tau)\le |\operatorname{supp}(\tau)|-1\le k-1.$ Since \(\dim \operatorname{aff}(\Delta(\Omega))=|\Omega|-1\ge k-1\), this affine subspace can be
extended to some \(L\in\mathcal L_{k-1}(\mu_0)\). Then
\(\operatorname{supp}(\tau)\subseteq L\), hence \(\operatorname{supp}(\tau)\subseteq K_L\). Therefore
\(\tau\in \mathcal I_k(L,\mu_0)\), so
\[
\E_\tau[\hat u^S(\mu)]
\le
W(L)
\le
\sup_{L'\in\mathcal L_{k-1}(\mu_0)} W(L').
\]
Taking the supremum over \(\tau\in\mathcal I(k,\mu_0)\) yields $V(k,\mu_0)\le \sup_{L\in\mathcal L_{k-1}(\mu_0)} W(L).$

Conversely, for every \(L\in\mathcal L_{k-1}(\mu_0)\), $\mathcal I_k(L,\mu_0)\subseteq \mathcal I(k,\mu_0),$ hence
\[
W(L)\le V(k,\mu_0).
\]
Taking the supremum over \(L\in\mathcal L_{k-1}(\mu_0)\) gives $\sup_{L\in\mathcal L_{k-1}(\mu_0)} W(L)\le V(k,\mu_0).$

\medskip
\noindent
\textbf{Step 2.} Fix \(L\in\mathcal L_{k-1}(\mu_0)\), and set $\hat u_L^S:=\hat u^S\big|_{K_L}.$ If \(\mu_0\) lies on the boundary of \(\Delta(\Omega)\), replace \(\Omega\) by
\(\operatorname{supp}(\mu_0)\). Bayes plausibility implies that every feasible posterior assigns zero
probability to \(\Omega\setminus \operatorname{supp}(\mu_0)\), so this does not change the problem.
By Proposition 1 of \cite{KG11}, Bayes plausibility is the only restriction on posterior
distributions. Therefore the fixed-slice problem can be written as
\[
W(L)
=
\sup\Bigl\{
\E_\tau[\hat u_L^S(\mu)]
\;\Bigm|\;
\tau\in \Delta(K_L),\ \E_\tau[\mu]=\mu_0,\ |\operatorname{supp}(\tau)|\le k
\Bigr\}.
\]

Define, for each \(x\in K_L\),
\[
\Gamma_L(x):=
\sup\Bigl\{
\sum_{i=1}^m \lambda_i \hat u_L^S(\mu_i)
\;\Bigm|\;
m\ge 1,\
\mu_i\in K_L,\
\lambda_i\ge 0,\
\sum_{i=1}^m\lambda_i=1,\
\sum_{i=1}^m\lambda_i\mu_i=x
\Bigr\}.
\]
By the standard representation of the concave envelope on a convex domain, the concavification result, $\Gamma_L=\operatorname{cav}(\hat u_L^S)$ on $K_L$. Since \(K_L\) has affine dimension at most \(k-1\), Carathéodory’s theorem implies that every
feasible decomposition in the definition of \(\Gamma_L(\mu_0)\) can be reduced to one with at most
\(k\) points. Hence
\[
\operatorname{cav}(\hat u_L^S)(
\mu_0)= \Gamma_L(\mu_0)
=
\sup\Bigl\{
\E_\tau[\hat u_L^S(\mu)]
\;\Bigm|\;
\tau\in \Delta(K_L),\ \E_\tau[\mu]=\mu_0,\ |\operatorname{supp}(\tau)|\le k
\Bigr\}
=
W(L).
\]

\medskip
\noindent
\textbf{Step 3.} Combining the conclusions of Step 1 and Step 2, we obtain that
\[
V(k,\mu_0)
=
\sup_{L\in\mathcal L_{k-1}(\mu_0)} W(L)
=
\sup_{L\in\mathcal L_{k-1}(\mu_0)}
\Bigl(\operatorname{cav}(\hat u_L^S)\Bigr)(\mu_0),
\]
This proves the proposition.

\end{proof}

\subsection{Value of An Additional Message}

\begin{proof}[\textbf{Proof of Theorem \ref{theorem:lowerbound}.}]
Let $\tau_{k+1}^*\in  \mathcal I(k+1,\mu_0)$ be optimal. If $|\supp(\tau_{k+1}^*)|\le k$, then $\tau_{k+1}^*\in \mathcal I(k,\mu_0)$, hence
$V(k+1,\mu_0)=V(k,\mu_0)$ and the claim is immediate. So suppose $\supp(\tau_{k+1}^*)=\{\mu_1,\dots,\mu_{k+1}\},$ with indices stated modulo $k+1$.

For each $i=1,\dots,k+1$, define
\[
\mu_{i,i+1}:=
\frac{\tau_{k+1}^*(\mu_i)}{\tau_{k+1}^*(\mu_i)+\tau_{k+1}^*(\mu_{i+1})}\mu_i
+
\frac{\tau_{k+1}^*(\mu_{i+1})}{\tau_{k+1}^*(\mu_i)+\tau_{k+1}^*(\mu_{i+1})}\mu_{i+1}.
\]
Let $\tau_{i,i+1}\in \mathcal I(k,\mu_0)$ be obtained from $\tau_{k+1}^*$ by replacing
$\mu_i$ and $\mu_{i+1}$ with $\mu_{i,i+1}$ and assigning to $\mu_{i,i+1}$ the
probability $\tau_{k+1}^*(\mu_i)+\tau_{k+1}^*(\mu_{i+1})$. This preserves Bayes
plausibility. Hence, by optimality of $V(k,\mu_0)$,
\[
V(k,\mu_0)\ge \E_{\tau_{i,i+1}}\!\left[\hat u^S(\mu)\right]
\qquad\text{for every } i=1,\dots,k+1.
\]
Averaging over $i$,
\[
V(k,\mu_0)\ge \frac{1}{k+1}\sum_{i=1}^{k+1}
\E_{\tau_{i,i+1}}\!\left[\hat u^S(\mu)\right].
\]

Now each term $\tau_{k+1}^*(\mu_j)\hat u^S(\mu_j)$ appears in exactly $k-1$ of
the $k+1$ expectations on the right-hand side, since $\mu_j$ is merged in exactly
two of the information structures $\tau_{i,i+1}$. Therefore,
\[
\frac{1}{k+1}\sum_{i=1}^{k+1}\E_{\tau_{i,i+1}}\!\left[\hat u^S(\mu)\right]
=
\frac{k-1}{k+1}V(k+1,\mu_0)
+\frac{1}{k+1}\sum_{i=1}^{k+1}
\bigl(\tau_{k+1}^*(\mu_i)+\tau_{k+1}^*(\mu_{i+1})\bigr)\hat u^S(\mu_{i,i+1}).
\]
Hence, if $\underline{\hat u^S}:=\inf_{\mu\in\Delta(\Omega)}\hat u^S(\mu),$ then
\[
V(k,\mu_0)\ge \frac{k-1}{k+1}V(k+1,\mu_0)+\frac{2}{k+1}\underline{\hat u^S},
\]
since  $\sum_{i=1}^{k+1}\bigl(\tau_{k+1}^*(\mu_i)+\tau_{k+1}^*(\mu_{i+1})\bigr)=2.$
Equivalently,
\[
V(k,\mu_0)-\underline{\hat u^S}
\ge
\frac{k-1}{k+1}\Bigl(V(k+1,\mu_0)-\underline{\hat u^S}\Bigr).
\]
Rearranging we get,
\[
\Delta_{k+1}
\le
\frac{2}{k+1}\Bigl(V(k+1,\mu_0)-\underline{\hat u^S}\Bigr)
\le
\frac{2}{k-1}\Bigl(V(k,\mu_0)-\underline{\hat u^S}\Bigr).
\]
Under the normalization used in the theorem, $\underline{\hat u^S}=0$, so this
reduces to
\[
\Delta_{k+1}\le \frac{2}{k+1}V(k+1,\mu_0)\le \frac{2}{k-1}V(k,\mu_0).
\]
This proves the theorem.
\end{proof}

\begin{proof}[\textbf{Proof of Corollary \ref{cor:scaling}.}]
For each integer $r\ge 2$, Theorem \ref{theorem:lowerbound} gives
\[
\Delta_{r+1}=V(r+1,\mu_0)-V(r,\mu_0)\le \frac{2}{r+1}V(r+1,\mu_0).
\]
Hence
\[
V(r,\mu_0)\ge \Bigl(1-\frac{2}{r+1}\Bigr)V(r+1,\mu_0)
= \frac{r-1}{r+1}V(r+1,\mu_0), \quad \implies \quad
\frac{V(r,\mu_0)}{r(r-1)}\ge \frac{V(r+1,\mu_0)}{(r+1)r}.
\]
Thus the normalized sequence $W(r):=\frac{V(r,\mu_0)}{r(r-1)}$ is weakly decreasing in $r$. Therefore, for $2\le \ell\le k\le m$,
\[
\frac{V(\ell,\mu_0)}{\ell(\ell-1)}
\ge
\frac{V(k,\mu_0)}{k(k-1)}
\ge
\frac{V(m,\mu_0)}{m(m-1)} \quad \implies \quad
\frac{k(k-1)}{\ell(\ell-1)} V(\ell,\mu_0)\ \ge\ V(k,\mu_0)\ \ge\
\frac{k(k-1)}{m(m-1)} V(m,\mu_0).
\]
Rearranging we can also state,
\[
V(\ell,\mu_0)\ge \frac{\ell(\ell-1)}{k(k-1)}V(k,\mu_0).
\]
Hence
\[
0\le V(k,\mu_0)-V(\ell,\mu_0)
\le
V(k,\mu_0)-\frac{\ell(\ell-1)}{k(k-1)}V(k,\mu_0) =\Bigl(1-\frac{\ell(\ell-1)}{k(k-1)}\Bigr)V(k,\mu_0).
\]
Finally note that,
\[
\Bigl(1-\frac{\ell(\ell-1)}{k(k-1)}\Bigr)V(k,\mu_0)
\le
\Bigl(1-\frac{\ell(\ell-1)}{k(k-1)}\Bigr)\frac{k(k-1)}{\ell(\ell-1)}V(\ell,\mu_0)
=
\Bigl(\frac{k(k-1)}{\ell(\ell-1)}-1\Bigr)V(\ell,\mu_0).
\]
\end{proof}

\begin{proof}[\textbf{Proof of Corollary \ref{corolla:bound}.}]
    This is the specialization of Corollary~\ref{cor:scaling} with $\ell=2$ and $m=n$.
\end{proof}

\subsection{Belief-Threshold Games}

\begin{proof}[\textbf{Proof of Lemma \ref{lem:UB-total-final}}]
Fix any feasible information structure \(\{(\tau_i,\mu_i)\}_{i=1}^k\), and let
\(
r:=\sum_{i=1}^k \mathbf{1}\{\mu_i\in \cup_j R_j\}.
\)
For each \(\mu_i\in \cup_j R_j\), choose \(\omega_i\) such that \(\mu_i(\omega_i)\ge T\), and let $J:=\{\omega_i:\mu_i\in \cup_j R_j\}.$ Then, the accounting identity can be stated as:
\[
T \cdot \sum_{i=1}^k \tau_i \mathbf{1}\{\mu_i\in \cup_j R_j\}
\le
\sum_{i:\mu_i\in \cup_j R_j}\tau_i\mu_i(\omega_i)
\le
\sum_{\omega\in J}\sum_{m=1}^k \tau_m\mu_m(\omega)
=
\sum_{\omega\in J}\mu_0(\omega)
\le
S_{|J|}
\le
S_r.
\]

If \(S_k\ge T\), then trivially $\sum_{i=1}^k \tau_i \mathbf{1}\{\mu_i\in \cup_j R_j\}\le 1,$ so \(V(k,\mu_0)\le 1\).

Now suppose \(S_k<T\). If \(r=k\), then every message induces a risky action, so $\sum_{i=1}^k \tau_i \mathbf{1}\{\mu_i\in \cup_j R_j\}=1.$ The accounting identity then gives $T\le S_r=S_k<T,$ a contradiction. Hence \(r\le k-1\). Applying the accounting identity again, \(T\sum_{i=1}^k \tau_i \mathbf{1} \{\mu_i\in \cup_j R_j\}\le S_r\le S_{k-1}.\) Therefore every feasible information structure yields ex-ante payoff at most \(S_{k-1}/T\), and hence \( V(k,\mu_0)\le \frac{S_{k-1}}{T}. \) Combining the two cases proves the lemma.
\end{proof}

\begin{proof}[\textbf{Proof of Theorem \ref{threshold-prop}.}] The claim can be verified from the construction in the main text.
\end{proof}

\begin{proof}[\textbf{Proof of Proposition \ref{thm:mono}.}]
If $k^*\le 2$, the first claim is immediate, and the second does not apply. Suppose $k^*>2$. Since $k^*=\min\{k:S_k\ge T\},$ we have $S_1=p_{(1)}<T$. With one message the posterior is $\mu_0$, so $V(1,\mu_0)=0$. By Theorem \ref{threshold-prop}, $V(2,\mu_0)=\frac{S_1}{T},$ hence $\Delta_2=\frac{p_{(1)}}{T}.$

For every $k=3,\dots,k^*-1$, Theorem \ref{threshold-prop} gives
\[
\Delta_k
=
V(k,\mu_0)-V(k-1,\mu_0)
=
\frac{S_{k-1}-S_{k-2}}{T}
=
\frac{p_{(k-1)}}{T}.
\]
Therefore, $\Delta_k=\frac{p_{(k-1)}}{T}$ for every $k=2,\dots,k^*-1.$ Since \(p_{(1)}\ge \cdots \ge p_{(n)}\), the sequence $\{\Delta_k\}_{k=2}^{k^*-1}$ is weakly decreasing.

If $k^*\ge 3$, then
\[
\Delta_{k^*}
=
V(k^*,\mu_0)-V(k^*-1,\mu_0)
=
1-\frac{S_{k^*-2}}{T}, \quad  \text{ and } \quad \Delta_{k^*-1}
=
\frac{p_{(k^*-2)}}{T}.
\]
Hence
\[
\Delta_{k^*}<\Delta_{k^*-1}
\iff
1-\frac{S_{k^*-2}}{T}<\frac{p_{(k^*-2)}}{T}
\iff
T<S_{k^*-2}+p_{(k^*-2)}.
\]

Finally, if $k>k^*$, then $V(k,\mu_0)=V(k-1,\mu_0)=1$, so $\Delta_k=0$.
\end{proof}

\begin{proof}[\textbf{Proof of Corollary~\ref{corollary:tight}}]

Fix \(k\le n-2\), and consider a prior such that $p_{(1)}=\cdots=p_{(k+1)}=\frac{T}{k+1},$ and $\sum_{\ell>k+1} p_{(\ell)}=1-T,$ with each remaining tail probability weakly below \(T/(k+1)\). Then $S_k=\frac{kT}{k+1}<T$ and $S_{k+1}=T$.  So, \(k^\ast=k+1\). By Theorem~\ref{threshold-prop}, $V(k,\mu_0)=\frac{S_{k-1}}{T}=\frac{k-1}{k+1},$ and  $V(k+1,\mu_0)=1.$ Hence
\[
\Delta_{k+1}
=
V(k+1,\mu_0)-V(k,\mu_0)
=
1-\frac{k-1}{k+1}
=
\frac{2}{k+1}.
\]
Therefore
\[
\Delta_{k+1}
=
\frac{2}{k+1}V(k+1,\mu_0)
=
\frac{2}{k-1}V(k,\mu_0).
\]
Thus the constants in Theorem~\ref{theorem:lowerbound} are attained in the baseline
belief-threshold model and are therefore sharp.

The same logic extends to the broader partition-event version of the belief-threshold discussed in Appendix \ref{app:threshold-generality}. Suppose risky action \(a_i\) is induced whenever \(\mu(E_i)\ge T_i\), where
\(E_1,\dots,E_m\) partition \(\Omega\). If $\frac{\mu_0(E_1)}{T_1}
=
\cdots
=
\frac{\mu_0(E_{k+1})}{T_{k+1}}
=
\frac{1}{k+1},$
and any remaining ratios \(\mu_0(E_i)/T_i\) are weakly below \(\frac{1}{k+1}\), then
\[
V(k,\mu_0)=\frac{k-1}{k+1},
\qquad
V(k+1,\mu_0)=1,
\qquad
\Delta_{k+1}=\frac{2}{k+1}.
\]
Therefore,
\[
\Delta_{k+1}
=
\frac{2}{k+1}V(k+1,\mu_0)
=
\frac{2}{k-1}V(k,\mu_0).
\]
\end{proof}
\begin{proof}[\textbf{Proof of Proposition \ref{prop:majorization-new}.}] Since \(\mu_0\) majorizes \(\nu_0\), we have
\(S_\ell(\mu_0)\ge S_\ell(\nu_0)\) for every \(\ell\).

\emph{(1)} We show \(V(k,\mu_0)\ge V(k,\nu_0)\). If \(S_k(\nu_0)\ge T\), then \(V(k,\nu_0)=1\). Since
\(S_k(\mu_0)\ge S_k(\nu_0)\ge T\), we also have \(V(k,\mu_0)=1\).

If \(S_k(\nu_0)<T\le S_k(\mu_0)\), then \(V(k,\mu_0)=1\) and
\(
V(k,\nu_0)=\frac{S_{k-1}(\nu_0)}{T}\le \frac{S_k(\nu_0)}{T}<1.
\)
Hence \(V(k,\mu_0)>V(k,\nu_0)\).

If \(S_k(\mu_0)<T\), then both priors are in the partial-persuasion region, and
\[
V(k,\mu_0)-V(k,\nu_0)
=
\frac{S_{k-1}(\mu_0)-S_{k-1}(\nu_0)}{T}
\ge 0.
\]

Therefore \(V(k,\mu_0)\ge V(k,\nu_0)\).
Since \(V(k,\mu)\) depends only on the ordered masses \((p_{(1)}(\mu),\dots,p_{(n)}(\mu))\), it is symmetric and Schur-convex.

\emph{(2)} Let \(k^*:=k^*(\nu_0)\). Then \(S_{k^*}(\nu_0)\ge T\), so also
\(S_{k^*}(\mu_0)\ge S_{k^*}(\nu_0)\ge T\). By minimality of the cutoff, $k^*(\mu_0)\le k^* = k^*(\nu_0).$

\emph{(3)} Suppose \(k^*(\mu_0)=k^*(\nu_0)=k^*\), with \(k^*\ge 2\).
Then $\Delta_{k^*}(\mu)
=
1-\frac{S_{k^*-2}(\mu)}{T}.$ Since \(S_{k^*-2}(\mu_0)\ge S_{k^*-2}(\nu_0)\), we get
\[
\Delta_{k^*}(\mu_0)
=
1-\frac{S_{k^*-2}(\mu_0)}{T}
\le
1-\frac{S_{k^*-2}(\nu_0)}{T}
=
\Delta_{k^*}(\nu_0).
\]

\emph{(4)} If \(k<k^*(\mu_0)\), then by part (2) also \(k<k^*(\nu_0)\).
Hence both priors are before the cutoff, so $\Delta_k(\mu_0)=\frac{p_{(k-1)}(\mu_0)}{T},$ and $\Delta_k(\nu_0)=\frac{p_{(k-1)}(\nu_0)}{T}.$ Therefore the coordinatewise dominance assumption implies $\Delta_k(\mu_0)\ge \Delta_k(\nu_0).$
\end{proof}

\begin{proof}[\textbf{Proof of Proposition \ref{prop-thresh}.}]
For fixed prior \(\mu_0\), the value function is
\[
V(k,T)=
\begin{cases}
1, & T\le S_k,\\[4pt]
\dfrac{S_{k-1}}{T}, & T>S_k.
\end{cases}
\]
Hence \(V(k,T)\) is constant on \((0,S_k]\) and strictly decreasing on \( T \in (S_k,1)\).
Therefore \(V(k,\mu_0)\) is weakly decreasing in \(T\).

Next,
\[
\Delta_k(T)=V(k,T)-V(k-1,T)=
\begin{cases}
0, & T\le S_{k-1},\\[6pt]
\displaystyle 1-\frac{S_{k-2}}{T}, & S_{k-1}<T\le S_k,\\[10pt]
\displaystyle \frac{S_{k-1}-S_{k-2}}{T}
=
\frac{p_{(k-1)}}{T}, & T>S_k.
\end{cases}
\]
On \(T\in(S_{k-1},S_k)\),
\(
\frac{d}{dT}\Delta_k(T)=\frac{S_{k-2}}{T^2}\ge 0,
\)
so \(\Delta_k\) is weakly increasing there (strict if \(S_{k-2}>0\)). On \(T>S_k\),
\(
\frac{d}{dT}\Delta_k(T)=-\frac{p_{(k-1)}}{T^2}<0,
\)
so \(\Delta_k\) is strictly decreasing there.

Moreover,
\[
\Delta_k(S_k)
=
1-\frac{S_{k-2}}{S_k}
=
\frac{p_{(k-1)}+p_{(k)}}{S_k}
>
\frac{p_{(k-1)}}{S_k}
=
\lim_{T\downarrow S_k,\;T>S_k}\Delta_k(T),
\]
so \(\Delta_k\) has a downward jump at \(T=S_k\). Hence \(\Delta_k\) is strictly
decreasing on \([S_k,1)\).
\end{proof}

\section{Extensions} \label{appendix:ext}

This appendix provides the details behind the roadmap in Section~\ref{sec:threshold-extend}. We keep the baseline model in Section \ref{sec:threshold}  but generalize the preferences. The receiver's payoffs are
\[
u^R(a_0,\omega_j)=0,\qquad
u^R(a_i,\omega_j)=
\begin{cases}
\frac{1}{T_i}-1, & j=i,\\
-1, & j\neq i.
\end{cases}
\]
And the sender's payoffs are $u^S(a_0,\omega)=0,$ and $u^S(a_i,\omega)=v_i\ge 0$ for $i>0$.

For expositional simplicity, assume \(T_i\in(1/2,1)\) for every \(i\).\footnote{This restriction is
not essential. If some \(T_i\le 1/2\), the same basic feasibility logic can still be applied, but the
geometry is less clean because the persuasive regions for different risky actions need not be
disjoint: a posterior may satisfy \(\mu(\omega_i)\ge T_i\) and \(\mu(\omega_j)\ge T_j\) for more than
one \(i\neq j\). One must then keep track explicitly of these overlaps, and of tie-breaking among
simultaneously optimal risky actions, so the statements become less transparent. We impose
\(T_i>1/2\) to avoid these case distinctions.} This restriction preserves the same geometry as in the
baseline model. Especially, it implies that once one risky action clears its threshold, no other risky action can do so.\footnote{Indeed,
if \(\mu(\omega_i)\ge T_i\), then action \(a_i\) yields expected payoff
\(\mu(\omega_i)/T_i-1\ge 0\). For every \(j\neq i\), $\mu(\omega_j)\le 1-\mu(\omega_i)\le 1-T_i<\frac12<T_j,$
so action \(a_j\) yields expected payoff \(\mu(\omega_j)/T_j-1<0\). Hence \(a_i\) is the only risky
best response at \(\mu\). If \(\mu(\omega_i)=T_i\), it may tie with \(a_0\), in which case ties are
broken in favor of the sender.}

Recall that the action-specific persuasion capacity is $c_i:=\frac{p_i}{T_i}.$ If action \(a_i\) is induced with ex-ante probability \(\tau_i\), then the persuasive messages
recommending \(a_i\) must together devote at least \(T_i \cdot \tau_i\) units of posterior mass to state
\(\omega_i\). Bayes plausibility limits that mass to \(p_i\), so \(\tau_i\le c_i\). Thus \(c_i\) is the
largest ex-ante frequency with which \(a_i\) can be induced. We  denote the ordered capacities by \(c_{(1)}\ge c_{(2)}\ge \cdots \ge c_{(n)}\) and define $C_\ell:=\sum_{i=1}^{\ell} c_{(i)}$ with  $C_0:=0.$

All the results presented below are constructive, similar to those in the baseline model. In Section \ref{sub-proofs}, we provide explicit Bayes plausible information structures (a full-persuasion construction and a construction with a residual message that achieve the stated values in the relevant regimes) and proofs of the results.

\subsection{Heterogeneous belief thresholds.} \label{sub-het-belief}

Suppose first that \(v_i=1\) for all \(i>0\), as in the baseline model. We can restate the baseline result (and construction) with the common ratio \(p_i/T\) replaced by the
action-specific capacity \(c_i=p_i/T_i\). The sender allocates scarce messages to the actions with
the largest capacities, so priors and thresholds matter only through the ordered vector
\((c_{(1)},\dots,c_{(n)})\). We formalize this in the result below.

\begin{Proposition}\label{prop:hetero-thresholds}
The sender's optimal value with at most \(k\) messages is
\[
V_{\mathrm{het}}(k,\mu_0)=
\begin{cases}
1, & C_k\ge 1,\\[4pt]
C_{k-1}, & C_k<1.
\end{cases}
\]
\end{Proposition}

Let \(\Delta_k:=V_{\mathrm{het}}(k,\mu_0)-V_{\mathrm{het}}(k-1,\mu_0)\) and redefine
\(k^\ast:=\min\{\ell:C_\ell\ge 1\}\). Before the cutoff, the \(k\)th message contributes the next-largest capacity, thus  $\Delta_k=c_{(k-1)}$ whenever $2\le k<k^\ast$. Beyond the cutoff, it has no value, thus $\Delta_k =0$ whenever $ k>k^\ast.$ Finally, at the cutoff, it eliminates the remaining default probability, so when $k=k^\ast$, there is a last-mile jump, and $\Delta_{k^*}=1-C_{k^\ast-2}.$

Comparative statics from the baseline model directly extend by replacing \((S_1,\dots,S_n)\) with partial sums \((C_1,\dots,C_n)\). Any change in priors or thresholds that weakly raises these partial
sums weakly raises \(V_{\mathrm{het}}(k,\mu_0)\) for every \(k\), weakly lowers \(k^\ast\), and,
conditional on leaving \(k^\ast\) unchanged, weakly lowers the last-mile increment
\(1-C_{k^\ast-2}\).

\subsection{A valued subset of risky actions.} \label{sub-valuedd}

We now go back to a common threshold \(T_i=T\), but instead suppose the sender benefits only from a subset
\(G\subseteq \{1,\dots,n\}\) of risky actions:
\[
u^S(a_i,\omega)=\mathbf{1}\{i\in G\},\qquad
u^S(a_0,\omega)=0.
\]
Let \(V_G(k,\mu_0)\) denote the sender's value. Write
\(p^G_{(1)}\ge p^G_{(2)}\ge \cdots\) for the prior masses
\(\{p_i:i\in G\}\) arranged in decreasing order, and define $S^G_\ell:=\sum_{j=1}^{\ell} p^G_{(j)}$ with  $S^G_0:=0.$

Only valuable states enter the relevant cumulative mass. Relative to the baseline model,
the extension simply replaces the full ordered prior profile by the ordered profile of valuable
masses.  We formalize this in the statement below.\footnote{This extension can also be viewed as a boundary-face version of the baseline threshold problem.
From the sender's perspective, all mass on states outside \(G\) is payoff-irrelevant except through
Bayes plausibility, so those states can be treated as a residual tail attached to the face spanned
by the valuable coordinates. Equivalently, the valued-subset problem behaves like a threshold
game on a reduced simplex whose only payoff-relevant risky directions are the coordinates in
\(G\), with the remaining prior mass pooled into a single residual state that supports the default.
This is why the value formula depends only on the ordered masses \((p^G_{(1)},p^G_{(2)},\dots)\).}

\begin{Proposition}\label{prop:valued-subset}
Assume \(T\in[1/2,1)\). For every \(k\ge 2\),
\[
V_G(k,\mu_0)= \begin{cases}
1, & S^G_k \ge T,\\[6pt]
\displaystyle \frac{S^G_{k-1}}{T}, & S^G_k < T.
\end{cases}
\]
\end{Proposition}

In particular, if \(G\subseteq G'\), then \(V_G(k,\mu_0)\le V_{G'}(k,\mu_0)\) for every \(k\),
so enlarging the valuable set weakly raises the sender's value and weakly lowers the number of
messages needed to eliminate the zero-payoff outcome.

We define $k_G^\ast:=\min\{\ell:S^G_\ell\ge T\},$
with the convention \(k_G^\ast=\infty\) if \(S^G_\ell<T\) for every \(\ell\). The marginal effect of adding a new valuable action \(j\notin G\) has two distinct components. If the new action does not change the cutoff, the gain is just the replacement effect in the top \(k-1\) valuable masses. If instead adding \(j\) lowers the cutoff enough that full value becomes attainable with \(k\) messages, the gain is the discrete last-mile jump.
\[
V_{G'}(k,\mu_0)-V_G(k,\mu_0)=
\begin{cases}
0, & k\ge k_G^\ast,\\[6pt]
\displaystyle 1-\frac{S^G_{k-1}}{T}, & k<k_G^\ast \ \text{and}\ k\ge k_{G'}^\ast,\\[10pt]
\displaystyle \frac{S^{G'}_{k-1}-S^G_{k-1}}{T}, & k<k_{G'}^\ast.
\end{cases}
\]
All comparative statics from the baseline belief-threshold model extend verbatim after replacing
\((p_{(1)},p_{(2)},\dots)\) and \((S_1,S_2,\dots)\) by \((p^G_{(1)},p^G_{(2)},\dots)\) and
\((S^G_1,S^G_2,\dots)\). In particular, a mean-preserving increase in concentration among the
valuable prior masses weakly raises \(V_G(k,\mu_0)\) for every \(k\), weakly lowers \(k_G^\ast\),
and, conditional on leaving the cutoff unchanged, weakly lowers the last-mile jump. Likewise,
a higher threshold \(T\) weakly lowers value and weakly raises the cutoff. The value of the
\(k\)th message is
\[
\Delta_k^G:=V_G(k,\mu_0)-V_G(k-1,\mu_0)=
\begin{cases}
0, & k>k_G^\ast,\\[4pt]
\displaystyle 1-\frac{S^G_{k_G^\ast-2}}{T}, & k=k_G^\ast,\\[10pt]
\displaystyle \frac{p^G_{(k-1)}}{T}, & 2\le k<k_G^\ast.
\end{cases}
\]
So, exactly as in the baseline model, marginal values are weakly decreasing before the cutoff,
there may be a discrete jump at the cutoff, and additional messages are valueless beyond it.

\subsection{Heterogeneous sender values.} \label{sub:het-val}

We now return to the full model with arbitrary thresholds \(T_i\) and sender values
\(v_i\ge 0\). Define the value-adjusted capacities $z_i:=v_i c_i = v_i\frac{p_i}{T_i},$ and let \(z_{(1)}\ge z_{(2)}\ge \cdots \ge z_{(n)}\) denote the ordered values. Finally, define $Z_\ell:=\sum_{j=1}^{\ell} z_{(j)}$ with $Z_0:=0$.

When \(C_k<1\), the default cannot be eliminated with \(k\) messages. In this case the
sender's problem can again be solved explicitly. Tractability here comes from the fact that the information-design problem collapses to a simple capacity-allocation problem. Since the default cannot be eliminated, at most \(k-1\) messages can be used to induce non-default actions. Moreover, the same accounting identity holds: action \(i\) can be induced with probability at most \(c_i\). Thus each action contributes at most \(v_i c_i\) to the sender's payoff, independently of the others. The sender's problem therefore becomes separable: choose up to \(k-1\) actions and allocate persuasive messages to the actions with the largest value-capacity products \(z_i\). This is what makes the problem explicit and yields \(V(k,\mu_0)=Z_{k-1}\).

\begin{Proposition}\label{prop:hetero-values-binding}
Assume \(C_k<1\). Then $V(k,\mu_0)=Z_{k-1}.$
\end{Proposition}

Once \(C_k\ge 1\), by contrast, the default can in principle be eliminated, and the sender's
choice is no longer separable across actions. Let \(x_i\) denote the ex-ante frequency with which
action \(a_i\) is induced. The sender then faces the allocation problem
\[
\max_{x_i\ge 0}\sum_i v_i x_i
\qquad \text{subject to} \qquad
x_i\le c_i \ \ \forall i,\qquad
\sum_i x_i = 1,\qquad
\bigl|\{i:x_i>0\}\bigr|\le k.
\]
The key difference from the \(C_k<1\) case is that the unit-mass constraint
\(\sum_i x_i=1\) now binds. Increasing the frequency of one risky action necessarily requires
reducing the frequency of another, so the sender can no longer evaluate actions independently by
their stand-alone contribution \(z_i\). An action may have a large \(z_i\) because it has
large capacity \(c_i\), even if its per-unit value \(v_i\) is modest; conversely, an action with a
small capacity but a high value may be attractive because it should be used first and exhausted
before lower-value actions are filled. Thus the sender must jointly choose a support of at most
\(k\) actions whose capacities together can absorb the full unit mass, and then determine how the
residual mass is split across the selected actions. This makes the problem a knapsack-type
selection problem with a cardinality constraint, rather than a simple ranking by \(z_i\).\footnote{
For example, let \(k=2\) and suppose
\((v_1,c_1)=(1,0.9)\), \((v_2,c_2)=(1,0.8)\), and \((v_3,c_3)=(3,0.2)\). Then
\((z_1,z_2,z_3)=(0.9,0.8,0.6)\), so ranking actions by \(z_i\) would favor \(1\) and \(2\).
But the optimal allocation is \(x_3=0.2\) and \(x_1=0.8\), which yields value \(1.4\), whereas
using actions \(1\) and \(2\) yields only \(1.0\). Once \(\sum_i x_i=1\) binds, the ordering by
\(z_i\) is therefore no longer sufficient.}

Some comparative statics are still immediate. The value \(V(k,\mu_0)\) is weakly increasing in
every \(v_i\) and every \(c_i\), since higher values raise the objective and higher capacities relax
the feasible set. But the effect of a parameter change is now support-dependent.  The value of an additional message is correspondingly support-dependent once \(C_k\ge 1\). Writing
\(\Delta_{k+1}:=V(k+1,\mu_0)-V(k,\mu_0)\), we still have \(\Delta_{k+1}\ge 0\), but the gain is no
longer pinned down by a simple order statistic such as \(z_{(k)}\). An additional message is
valuable through support expansion: it permits the sender to bring one more action into the
candidate set and then rebalance probability mass across the selected actions.

If \(\bar k\) denotes the smallest support size of an
optimal solution to the problem without the cardinality constraint, then
\(\Delta_{k+1}=0\) for all \(k\ge \bar k\). In general \(\bar k\) can exceed the smallest \(k\) for
which \(C_k\ge 1\), so the ability to eliminate the default does not, by itself, exhaust the value
of additional messages.

A particularly tractable special case arises when higher sender values are aligned with larger
persuasion capacities. In that case the support choice is no longer combinatorial, and a greedy
allocation is optimal. Suppose the actions are indexed so that $v_1\ge v_2\ge \cdots \ge v_n$ and $c_1\ge c_2\ge \cdots \ge c_n.$  Also recall that $k^\ast:=\min\{\ell:C_\ell\ge 1\}.$ Then
\[
V(k,\mu_0)=
\begin{cases}
\displaystyle Z_{k-1}, & k<k^\ast,\\[10pt]
\displaystyle Z_{k^*-1}
+ v_{k^\ast}\bigl(1-C_{k^\ast-1}\bigr), & k\ge k^\ast.
\end{cases}
\]
Under alignment, the sender first exhausts the highest-value, highest-capacity actions and then,
if necessary, uses one final action to absorb the remaining probability mass. Before the cutoff,
the \(k\)th message is worth exactly \(z_{k-1}\). At the cutoff, the value is \(Z_{k^\ast-1}+v_{k^\ast}(1-C_{k^\ast-1})\); when \(k^\ast>1\), this means the cutoff message adds \(z_{k^\ast-1}+v_{k^\ast}(1-C_{k^\ast-1})\) relative to \(k^\ast-1\) messages. Once \(k>k^\ast\), additional messages are valueless.

\subsection{Proofs of statements in Extensions} \label{sub-proofs}

Write \(p_i:=\mu_0(\omega_i)\) and \(c_i:=p_i/T_i\) for each \(i\). Set \(C_0:=0\) and
\(S_0^G:=0\). When an ordered partial sum is indexed beyond the end of the list, interpret it as
the full sum.

A basic upper bound will be used repeatedly. If an experiment induces action \(a_i\) with total
ex-ante probability \(s_i\), then $T_i s_i\le p_i,$ hence  $s_i\le c_i.$
Indeed, every message that induces \(a_i\) must satisfy \(\mu(\omega_i)\ge T_i\), so summing
\(\tau(\mu)\mu(\omega_i)\ge T_i\tau(\mu)\) over all such messages and using Bayes plausibility
yields \(p_i\ge T_i s_i\).

\medskip
\noindent
\textbf{Two constructions used repeatedly.} The proofs below use two explicit Bayes plausible experiments.

\medskip
\noindent
\underline{\emph{1. Full-persuasion construction.}}
Fix a nonempty set \(I\subseteq\{1,\dots,n\}\) and numbers \((\tau_i)_{i\in I}\) such that $0<\tau_i\le c_i$ for $i \in I$, and $\sum_{i\in I}\tau_i=1.$
Define
\[
b_j:=
\begin{cases}
p_j-T_j\tau_j,& j\in I,\\
p_j,& j\notin I,
\end{cases}
\qquad \text{and} \qquad
B:=\sum_{j=1}^n b_j
=
1-\sum_{i\in I}T_i\tau_i.
\]
Since \(\tau_i\le p_i/T_i\), we have \(b_j\ge 0\) for all \(j\). Also \(B>0\), because $B
=
1-\sum_{i\in I}T_i\tau_i
>
1-\sum_{i\in I}\tau_i
=0,$ where the strict inequality uses \(T_i<1\) for all \(i \in I\).

For each state \(j\) and each message \(i\in I\), define the ex-ante mass
\[
x_{ji}
:=
\mathbf 1_{\{j=i\}}\,T_i\tau_i
+
b_j\frac{(1-T_i)\tau_i}{B}.
\]
Then, for every state \(j\),
\[
\sum_{i\in I}x_{ji}
=
\mathbf 1_{\{j\in I\}}\,T_j\tau_j
+
b_j\frac{\sum_{i\in I}(1-T_i)\tau_i}{B}.
\]
Since \(\sum_{i\in I}\tau_i=1\), we have that $\sum_{i\in I}(1-T_i)\tau_i
=
1-\sum_{i\in I}T_i\tau_i
=
B,$ so \[\sum_{i\in I}x_{ji}= \mathbf 1_{\{j\in I\}}T_j\tau_j+b_j=p_j.\]
Likewise, for each message \(i\in I\),
\[
\sum_{j=1}^n x_{ji}
=
T_i\tau_i+\frac{(1-T_i)\tau_i}{B}\sum_{j=1}^n b_j
=
T_i\tau_i+(1-T_i)\tau_i
=
\tau_i.
\]
Hence \((x_{ji})\) defines a Bayes plausible experiment with message probabilities
\((\tau_i)_{i\in I}\).

Let \(\mu_i\) denote the posterior induced by message \(i\). Then
\[
\mu_i(\omega_i)=\frac{x_{ii}}{\tau_i}
=
T_i+\frac{b_i(1-T_i)}{B}
\ge T_i.
\]
Moreover, for every \(j\neq i\),
\[
\mu_i(\omega_j)\le 1-\mu_i(\omega_i)\le 1-T_i\le \frac12 \le T_j.
\]
Thus \(a_i\) is a receiver best response after message \(i\).

\medskip
\noindent
\underline{\emph{2. Partial persuasion construction with a residual message.}}
Fix \(I\subseteq\{1,\dots,n\}\) such that $\sum_{i\in I}c_i<1.$ Define $\tau_i:=c_i=\frac{p_i}{T_i},$ for $i\in I$. Moreover, $\tau_{\text{last}}:=1-\sum_{i\in I}c_i>0,$ and $P_I:=\sum_{i\in I}p_i.$ Since \(T_i<1\), we have \(c_i=p_i/T_i\ge p_i\), hence \(P_I\le \sum_{i\in I}c_i<1\).
For \(j\notin I\), let $\nu_j:=\frac{p_j}{1-P_I}.$

Define posteriors \((\mu_i)_{i\in I}\)  by $\mu_i(\omega_i)=T_i,$ and  $\mu_i(\omega_j)=0$ whenever $j\in I$ and $j\neq i$. When $j\notin I$, set $\mu_i(\omega_j)=(1-T_i)\nu_j\ \ (j\notin I),$

Define the residual posterior \(\mu_{\text{last}}\) by $\mu_{\text{last}}(\omega_j)=0$ whenever $j\in I$, and  $\mu_{\text{last}}(\omega_j)=\nu_j$ whenever $j\notin I$.

Each \(\mu_i\) is a probability distribution, because
\[
T_i+\sum_{j\notin I}(1-T_i)\nu_j
=
T_i+(1-T_i)\sum_{j\notin I}\nu_j
=
T_i+(1-T_i)=1.
\]
Also \(\mu_{\text{last}}\) is a probability distribution since \(\sum_{j\notin I}\nu_j=1\).

For each \(i\in I\), message \(i\) satisfies $\mu_i(\omega_i)=T_i,$ and $\mu_i(\omega_j)\le 1-T_i\le \frac12\le T_j$ for $j\neq i$. So, \(a_i\) is a receiver best response after message \(i\).

It remains to check Bayes plausibility.  For \(j\in I\), $\tau_i\mu_i(\omega_i)=c_iT_i=p_i$. Moreover, $\tau_\ell\mu_\ell(\omega_i)=0$ for $\ell \neq i$. Finally,  $\tau_{\text{last}}\mu_{\text{last}}(\omega_i)=0.$ Thus state \(\omega_i\) has total ex-ante mass \(p_i\). For \(j\notin I\),
\[
\sum_{i\in I}\tau_i\mu_i(\omega_j)+\tau_{\text{last}}\mu_{\text{last}}(\omega_j)
=
\sum_{i\in I}c_i(1-T_i)\nu_j+\tau_{\text{last}}\nu_j
=
\left(\sum_{i\in I}(c_i-p_i)+\tau_{\text{last}}\right)\nu_j.
\]
Since \(\tau_{\text{last}}=1-\sum_{i\in I}c_i\), $\sum_{i\in I}(c_i-p_i)+\tau_{\text{last}}
=
1-\sum_{i\in I}p_i
=
1-P_I.$ Therefore
\[
\sum_{i\in I}\tau_i\mu_i(\omega_j)+\tau_{\text{last}}\mu_{\text{last}}(\omega_j)
=
(1-P_I)\nu_j
=
p_j.
\]
So the experiment is Bayes plausible.

Finally, if \(\tau_{\text{last}}>c_j\) for every \(j\notin I\), then for \(j\notin I\),
\[
\mu_{\text{last}}(\omega_j)=\nu_j=\frac{p_j}{1-P_I}\le \frac{p_j}{\tau_{\text{last}}}<T_j,
\]
where the last inequality follows from \(\tau_{\text{last}}>c_j=p_j/T_j\), and the first inequality uses
\(1-P_I\ge \tau_{\text{last}}\). Hence the residual message induces the default action \(a_0\).

\begin{proof}[\textbf{Proof of Proposition~\ref{prop:hetero-thresholds}}]
Relabel actions so that $c_1\ge c_2\ge \cdots \ge c_n$ and  $C_\ell=\sum_{i=1}^{\ell}c_i$. Let \(s_i\) denote the total ex-ante probability with which action \(a_i\) is chosen in some
feasible experiment. By the preliminary bound, \(s_i\le c_i\) for every \(i\in\{1,\dots,n\}\).

Suppose first that \(C_k<1\). If every message induced a risky action, then the total probability
of risky actions would be \(1\). Let the targeted states be $I:=\{i:s_i>0\}.$ Since there are at most \(k\) messages, \(|I|\le k\). Hence
\[
1=\sum_{i\in I}s_i\le \sum_{i\in I}c_i\le C_k<1,
\]
a contradiction.  Therefore at least one message must be non-persuasive, so \(|I|\le k-1\). It
follows that
\[
V_{\mathrm{het}}(k,\mu_0)=\sum_{i=1}^n s_i
=
\sum_{i\in I}s_i
\le \sum_{i\in I}c_i
\le C_{k-1}.
\]

If we instead suppose \(C_k\ge 1\), the trivial bound \(V_{\mathrm{het}}(k,\mu_0)\le 1\) holds.

\medskip
We proceed by proving attainability in both cases. First consider the case \(C_k<1\). Take \(I=\{1,\dots,k-1\}\), set $\tau_i:=c_i$ for $i \in I$ and   $\tau_{\text{last}}:=1-C_{k-1}>0.$ and apply the partial persuasion construction.

For any \(j\notin I\), $c_j\le c_k<1-C_{k-1}=\tau_{\text{last}},$ since \(C_k=C_{k-1}+c_k<1\). Hence the residual message induces the default.

Each message \(i\in I\) induces some risky action, and every risky action has sender value \(1\). Therefore the
sender obtains $\sum_{i=1}^{k-1}\tau_i=C_{k-1}.$
We conclude \(V_{\mathrm{het}}(k,\mu_0)\ge C_{k-1}\), and hence $V_{\mathrm{het}}(k,\mu_0)=C_{k-1}$ whenever $C_k<1.$

Now consider the case \(C_k\ge 1\). Let $k^*:=\min\{\ell
:C_\ell\ge 1\}.$ Define $\tau_i:=c_i$ for $i<k^*$ and $\tau_{k^*}:=1-C_{k^*-1}.$ Then \(\tau_{k^*}\le c_{k^*}\) and \(\sum_{i=1}^{k^*}\tau_i=1\).

By the full-persuasion construction, these frequencies are implementable with at most \(k^*\le k\) persuasive messages and no residual message. Each such message induces a risky action, hence yields sender payoff \(1\). Therefore \(V_{\mathrm{het}}(k,\mu_0)=1\).

This proves
\[
V_{\textnormal{het}}(k,\mu_0)=
\begin{cases}
1, & C_k\ge 1,\\[4pt]
C_{k-1}, & C_k<1.
\end{cases}
=
\begin{cases}
1, & k\ge k^* .\\[4pt]
C_{k-1}, & k<k^*.
\end{cases}
\]

The marginal value formula follows directly:
\[
\Delta_k=
V_{\mathrm{het}}(k,\mu_0)-V_{\mathrm{het}}(k-1,\mu_0)
=
\begin{cases}
c_{(k-1)}, & 2\le k<k^*,\\[4pt]
1-C_{k^*-2}, & k=k^*,\\[4pt]
0, & k>k^*.
\end{cases}
\]
Finally, every comparative-static claim in the text follows immediately from the closed form above,
since \(V_{\mathrm{het}}(k,\mu_0)\) depends on \((\mu_0,T_i)\) only through the ordered partial sums
\((C_1,\dots,C_n)\).
\end{proof}

\begin{proof}[\textbf{Proof of Proposition~\ref{prop:valued-subset}.}]
For \(i\in G\), let \(s_i\) denote the total ex-ante probability with which \(a_i\) is chosen in some feasible experiment. Then $V_G(k,\mu_0)=\sum_{i\in G}s_i.$

Since the threshold is common and equal to \(T\), the preliminary bound gives $Ts_i\le p_i$ for every $i \in G$.

Suppose first that \(k<k_G^*\). Then \(S_k^G<T\). If all \(k\) messages induced valuable actions,
the total probability of valuable actions would be \(1\). Let $I:=\{i\in G:s_i>0\}.$ Then \(|I|\le k\), and therefore
\[
T
=
T\sum_{i\in G}s_i
=
T\sum_{i\in I}s_i
\le \sum_{i\in I}p_i
\le S_k^G
<
T,
\]
a contradiction. Hence at least one message must yield sender payoff \(0\), so \(|I|\le k-1\). It
follows that
\[
V_G(k,\mu_0)
=
\sum_{i\in I}s_i
\le
\frac{1}{T}\sum_{i\in I}p_i
\le
\frac{S_{k-1}^G}{T}.
\]

To attain this bound, let \(I\subseteq G\) be the set of valuable states contributing to
\(S_{k-1}^G\), i.e. the \(\min\{k-1,|G|\}\) valuable states with largest prior masses. Set
\[
\tau_i:=\frac{p_i}{T} \text{ for } \ i\in I,
\quad \text{ and } \quad
\tau_{\text{last}}:=1-\frac{S_{k-1}^G}{T}>0.
\]
Apply the partial persuasion construction with common threshold \(T_i=T\). For each \(i\in I\), message
\(i\) induces a valuable action and hence yields sender payoff \(1\).

Now take any \(j\in G\setminus I\). Since \(j\) does not belong to the top \(k-1\) valuable states, $p_j\le p_{(k)}^G,$ and $S_{k-1}^G+p_{(k)}^G=S_k^G<T.$ Thus $p_j<T-S_{k-1}^G=T\tau_{\text{last}},$
so
\[
\mu_{\text{last}}(\omega_j)=\nu_j=\frac{p_j}{1-P_I}\le \frac{p_j}{\tau_{\text{last}}}<T,
\]
where here \(P_I=\sum_{i\in I}p_i=S_{k-1}^G\).
Hence the residual message does not induce any valuable action. It may induce the default or an
unvalued risky action, but in either case its sender payoff is \(0\). Therefore
\[
V_G(k,\mu_0)
=
\sum_{i\in I}\tau_i
=
\frac{1}{T}\sum_{i\in I}p_i
=
\frac{S_{k-1}^G}{T}.
\]

\medskip
Now suppose \(k\ge k_G^*\). For notational simplicity let $k^*=k_G^*$ so that $S_{k^*}^G\ge T.$

Let \(I\subseteq G\) be the set of valuable states contributing to \(S_{k^*}^G\), and set $P_I:=\sum_{i\in I}p_i=S_{k^*}^G.$

Define $\tau_i:=\frac{p_i}{P_I}$ for all $i\in I$. By construction \(\sum_{i\in I}\tau_i=1\). Since \(P_I\ge T\),
\[
\tau_i=\frac{p_i}{P_I}\le \frac{p_i}{T}=c_i.
\]
Hence the full-persuasion construction applies. Every persuasive
message has some action in \(G\) as a receiver best response, and every action in \(G\) gives sender
value \(1\). So \(V_G(k,\mu_0)=1\).

This proves
\[
V_G(k,\mu_0)=
\begin{cases}
1, & k\ge k_G^\ast,\\[6pt]
\displaystyle \frac{S^G_{k-1}}{T}, & k<k_G^\ast.
\end{cases}
\]

The remaining claims follow immediately from the formula.
\end{proof}

\begin{proof}[\textbf{Proof of Proposition~\ref{prop:hetero-values-binding}.}]
Let \(s_i\) denote the total ex-ante probability with which action \(a_i\) is chosen in some
feasible experiment. Then $V(k,\mu_0)=\sum_{i=1}^n v_i s_i.$ By the preliminary bound, $s_i\le c_i$ for every $i$.

Assume \(C_k<1\). We first prove the upper bound. If every message yielded strictly positive sender
payoff, then all \(k\) messages would induce risky actions with positive sender value. Let $I:=\{i:v_i s_i>0\}.$ Then \(|I|\le k\), and the total probability of those actions would be \(1\). Therefore
\[
1=\sum_{i\in I}s_i\le \sum_{i\in I}c_i\le C_k<1,
\]
a contradiction. Hence at least one message must yield sender payoff \(0\), so \(|I|\le k-1\).
Consequently,
\[
V(k,\mu_0)
=
\sum_{i\in I}v_i s_i
\le
\sum_{i\in I}v_i c_i
\le
\sum_{r=1}^{k-1}z_{(r)},
\]
where \(z_i:=v_i c_i\) and \(z_{(1)}\ge \cdots \ge z_{(n)}\) is the decreasing rearrangement.

For the lower bound, let \(I\) be the set of indices corresponding to the \(k-1\) largest values of
\(z_i\). Set $\tau_i:=c_i,$ for $i\in I$ and $\tau_{\text{last}}:=1-\sum_{i\in I}c_i>0,$ and apply the partial persuasion construction. For each \(i\in I\), message \(i\) has \(a_i\) as a receiver
best response, so its sender payoff is at least \(v_i\).

We claim that the residual message yields sender payoff \(0\). Suppose instead that it yielded a
strictly positive sender payoff. Then it would induce some risky action \(a_j\) with \(v_j>0\) and
\(j\notin I\). In particular,
\[
\mu_{\text{last}}(\omega_j)=\nu_j=\frac{p_j}{1-P_I}\ge T_j.
\]
Hence
\[
c_j=\frac{p_j}{T_j}\ge 1-P_I\ge \tau_{\text{last}},
\]
so \(\sum_{i\in I}c_i+c_j\ge 1.\)

But \(|I|+1\le k\), and \(C_k\) is the sum of the \(k\) largest \(c\)-values. Therefore $C_k\ge \sum_{i\in I}c_i+c_j\ge 1,$ contradicting the hypothesis \(C_k<1\). So the residual message indeed yields sender payoff \(0\).

Therefore the sender obtains at least
\[
\sum_{i\in I}v_i\tau_i
=
\sum_{i\in I}v_i c_i
=
\sum_{r=1}^{k-1}z_{(r)}.
\]
Combining this with the upper bound yields
\[
V(k,\mu_0)=\sum_{r=1}^{k-1}z_{(r)}.
\]

The marginal value formula is immediate:
\[
\Delta_k
=
V(k,\mu_0)-V(k-1,\mu_0)
=
z_{(k-1)}.
\]
\end{proof}

\medskip
\noindent
\textbf{Claim about aligned values and capacities.} Assume now that actions are indexed so that $v_1\ge v_2\ge \cdots \ge v_n$ and $c_1\ge c_2\ge \cdots \ge c_n.$ Then \(z_i=v_i c_i\) is also weakly decreasing in \(i\).

If \(C_k<1\), Proposition~\ref{prop:hetero-values-binding} already gives $V(k,\mu_0)=\sum_{i=1}^{k-1}z_i.$

Suppose now that \(C_k\ge 1\), and let $k^*=\min\{r\le k:C_r\ge 1\}.$ For any feasible experiment, let \(s_i\) denote the total ex-ante probability with which action
\(a_i\) is chosen. Then \(0\le s_i\le c_i\) for all \(i\), and \(\sum_i s_i\le 1\). Therefore
\[
\sum_{i=1}^n v_i s_i
=
v_{k^*}\sum_{i=1}^n s_i
+
\sum_{i=1}^{k^*-1}(v_i-v_{k^*} )s_i
+
\sum_{i=k^*+1}^n (v_i-v_{k^*})s_i
\le
v_{k^*}+\sum_{i=1}^{k^*-1}(v_i-v_{k^*})c_i.
\]
That is,
\[
V(k,\mu_0)\le \sum_{i=1}^{{k^*}-1}z_i+v_{k^*}\bigl(1-C_{k^*-1}\bigr).
\]

 To attain this bound, define $\tau_i:=c_i$ for $i<k^*$ and $\tau_{k^*}:=1-C_{k^*-1}$. Finally, $\tau_i:=0$ for $i > k^*$.

 Since \(C_{k^*} \ge 1\) and \(C_{k^*-1}<1\), we have \(0<\tau_{k^*}\le c_{k^*}\), and
\(\sum_{i=1}^{k^*}\tau_i=1\). Apply the full-persuasion construction on \(I=\{1,\dots,k^*\}\).

For \(j<k^*\), $\tau_j=c_j=\frac{p_j}{T_j}$. So, $b_j=p_j-T_j\tau_j=0.$

Hence, for any \(i<k^*\), $\mu_i(\omega_i)=T_i,$ and $\mu_i(\omega_j)=0$ for $j<k^*$ and $j\neq i$. Thus no action with index \(j<i\) can be a receiver best response after message \(i\). Any action
with index \(j>i\) has sender value at most \(v_i\). Since \(a_i\) itself is a receiver best
response, sender-favored tie-breaking yields sender payoff exactly \(v_i\) on message \(i\).

For message \(k^*\), we likewise have $\mu_{k^*}(\omega_j)=0$ for $j<k^*$, and $\mu_{k^*}(\omega_{k^*})\ge T_{k^*}.$
So no higher-value action can be a receiver best response after message \(k^*\), whereas \(a_{k^*}\) is.
Thus message \(k^*\) yields sender payoff exactly \(v_{k^*}\).

Therefore this experiment attains
\[
\sum_{i=1}^{k^*-1}v_i c_i+v_{k^*}\bigl(1-C_{k^*-1}\bigr)
=
\sum_{i=1}^{k^*-1}z_i+v_{k^*}\bigl(1-C_{k^*-1}\bigr).
\]
Combined with the upper bound, this gives
\[
V(k,\mu_0)=
\begin{cases}
\displaystyle \sum_{i=1}^{k-1} z_i, & C_k<1,\\[10pt]
\displaystyle \sum_{i=1}^{k^*-1} z_i
+ v_{k^*}\bigl(1-C_{k^*-1}\bigr), & C_k\ge 1.
\end{cases}
\]

The marginal-value statements follow immediately. \qed

\section{Cheap Talk with Transparent Motives}\label{cheaptalk}

\medskip
\noindent
\textbf{The Model.} \cite{lipnowski2020cheap} study a cheap-talk model with state-independent sender preferences. Relative to the persuasion environment in the main text, three features differ. First, the sender cannot commit ex ante to an information structure. Second, the sender's payoff depends only on the receiver's action, so \(u^S:A\to \mathbb{R}\). Third, \cite{lipnowski2020cheap} focus on the rich-message case \(|M|\geq |\Omega|\). In this appendix, we keep the first two features and replace the third with a coarse message space \(|M|\leq k<|\Omega|\).

A perfect Bayesian equilibrium consists of a messaging strategy \(\pi:\Omega\to\Delta(M)\), a receiver strategy \(\hat a:M\to\Delta(A)\), and a belief system \(\mu(\cdot\mid m)\in\Delta(\Omega)\) such that:
\begin{enumerate}
    \item For every on-path message \(m\), \(\mu(\cdot\mid m)\) is obtained from \(\mu_0\) and \(\pi\) by Bayes' rule;
    \item For every \(m\in M\), \(\hat a(m)\) is supported on
    \[
    BR(\mu(\cdot\mid m))
    :=
    \arg\max_{a\in A}\int_\Omega u^R(a,\omega)\,d\mu(\omega\mid m);
    \]
    \item For every \(\omega\in\Omega\), \(\pi(\omega)\) is supported on
    \[
    \arg\max_{m\in M}\int_A u^S(a)\,d\hat a(a\mid m).
    \]
\end{enumerate}

\medskip
\noindent
\textbf{Belief Based Approach.} As in \cite{lipnowski2020cheap}, it is convenient to work with the ex-ante distribution of receiver posteriors. Given an equilibrium \((\pi,\hat a,\mu)\), let \(\tau_\pi\in\Delta(\Delta(\Omega))\) denote the induced distribution of posteriors. Bayes plausibility implies $\E_{\tau_\pi}[\mu]=\mu_0.$

Since the sender has at most \(k\) messages, \(|\operatorname{supp}(\tau_\pi)|\leq k\). Thus the feasible set of posterior distributions is
\[ \mathcal
I(k,\mu_0)
:=
\left\{
\tau\in\Delta(\Delta(\Omega))
:
\E_\tau[\mu]=\mu_0,\;
|\operatorname{supp}(\tau)|\leq k
\right\}.
\]

To connect with the optimal-compression result in the main text, for each \(L\in \mathcal L_{k-1}(\mu_0)\) define $K_L:=\Delta(\Omega)\cap L$ and the feasible set of posterior distributions on the compression:
\[
\mathcal I_k(L,\mu_0)
:=
\left\{
\tau\in\Delta(\Delta(\Omega))
:
\E_\tau[\mu]=\mu_0,\;
\operatorname{supp}(\tau)\subseteq K_L,\;
|\operatorname{supp}(\tau)|\leq k
\right\}.
\]

\medskip
\noindent
\textbf{Sender Payoff.} For each posterior \(\mu\in\Delta(\Omega)\), let
\[
BR(\mu):=\arg\max_{a\in A}\int_\Omega u^R(a,\omega)\,d\mu(\omega), \quad \text{and} \quad V(\mu):=\operatorname{co}\{u^S(a):a\in BR(\mu)\}\subset\mathbb{R}.
\]
Also define the Sender's interim value function
\[
\hat u^S(\mu):=\max V(\mu)=\max_{a\in BR(\mu)}u^S(a).
\]
Here \(V(\mu)\) is the set of sender continuation values compatible with receiver best responses at posterior \(\mu\), while \(\hat u^S(\mu)\) is the sender-preferred continuation value at \(\mu\).

By \cite{aumann2003long} and \cite{lipnowski2020cheap}, in the rich-message case an outcome pair \((\tau,z)\) is an equilibrium outcome if and only if \(\tau\) is Bayes plausible and
\[
z\in\bigcap_{\mu\in\operatorname{supp}(\tau)}V(\mu).
\]
The same argument applies for coarse communication when the unrestricted Bayes plausible set is replaced by \(\mathcal I(k,\mu_0)\).

\begin{Lemma}\label{lipravidextend}
An outcome pair \((\tau,z)\) is a cheap-talk equilibrium outcome with \(|M|\leq k\) if and only if $\tau\in \mathcal  I(k,\mu_0)$ and $z\in\bigcap_{\mu\in\operatorname{supp}(\tau)}V(\mu).$
\end{Lemma}

\medskip
\noindent
\textbf{Securability and Quasi-concavification.} Following \cite{lipnowski2020cheap}, say that \(\tau\in \mathcal  I(k,\mu_0)\) \emph{\(k\)-secures} a payoff \(z\) if $\hat u^S(\mu)\geq z$ for every $\mu\in\operatorname{supp}(\tau).$

Equivalently, \(z\) is \emph{\(k\)-securable} if there exists some \(\tau\in\mathcal  I(k,\mu_0)\) that \(k\)-secures \(z\). By Lemma \ref{lipravidextend} and the argument in \cite{lipnowski2020cheap}, a payoff \(z\) is attainable in equilibrium with \(|M|\leq k\) if and only if it is \(k\)-securable. Hence the sender-preferred equilibrium payoff is
\[
V^{CT}(k,\mu_0)
:=
\max_{\tau\in \mathcal  I(k,\mu_0)}
\min_{\mu\in\operatorname{supp}(\tau)}
\hat u^S(\mu).
\]

The compression result from the main text implies that every \(\tau\in \mathcal  I(k,\mu_0)\) is supported on some slice \(K_L\) with \(L\in \mathcal L_{k-1}(\mu_0)\). Thus coarse cheap talk can be viewed as first choosing a \((k-1)\)-dimensional compression \(L\), and then solving the transparent-motives problem on \(K_L\).

For each \(L\in \mathcal L_{k-1}(\mu_0)\), let \(\bar u^S_L\) denote the upper-semicontinuous quasiconcave envelope of \(\hat u^S|_{K_L}\) on \(K_L\).

\begin{Proposition}\label{k-qcav}
For every \(k\geq 2\),
\[
V^{CT}(k,\mu_0)
=
\sup_{L\in \mathcal L_{k-1}(\mu_0)}
\left(
\sup_{\tau\in \mathcal I_k(L,\mu_0)}
\inf_{\mu\in\operatorname{supp}(\tau)}
\hat u^S(\mu)
\right)
=
\sup_{L\in \mathcal L_{k-1}(\mu_0)}
\bar u^S_L(\mu_0).
\]
\end{Proposition}

\begin{proof}[\textbf{Proof of Proposition \ref{k-qcav}.}]
By the same compression argument as in Proposition \ref{Optimal Summary}, any \(\tau\in \mathcal  I(k,\mu_0)\) is supported on some \(K_L\) with \(L\in \mathcal L_{k-1}(\mu_0)\), and conversely \( \mathcal  I_k(L,\mu_0)\subseteq \mathcal  I(k,\mu_0)\) for every such \(L\). Therefore
\[
V^{CT}(k,\mu_0)
=
\sup_{L\in \mathcal L_{k-1}(\mu_0)}
\left(
\sup_{\tau\in \mathcal I_k(L,\mu_0)}
\inf_{\mu\in\operatorname{supp}(\tau)}
\hat u^S(\mu)
\right).
\]

Fix \(L\in \mathcal L_{k-1}(\mu_0)\). Since \(K_L\) has affine dimension at most \(k-1\), Carath\'eodory's theorem implies that whenever $\mu_0\in\operatorname{co}\{\mu\in K_L:\hat u^S(\mu)\geq z\},$ the same inclusion can be characterized using at most \(k\) points. Hence, once the slice \(K_L\) is fixed, the cardinality bound \(|M|\leq k\) does not impose any additional restriction beyond Bayes plausibility.

Applying \cite{lipnowski2020cheap} on the compressed domain \(K_L\), the sender's best equilibrium payoff on that slice is exactly the quasiconcave envelope of \(\hat u^S|_{K_L}\) evaluated at \(\mu_0\), namely \(\bar u^S_L(\mu_0)\). Taking the supremum over \(L\in \mathcal L_{k-1}(\mu_0)\) yields the result.
\end{proof}

Proposition \ref{k-qcav} is the cheap-talk analogue of the compress--then--concavify representation in the main text. With coarse communication, the sender first chooses an optimal compression \(L\in \mathcal L_{k-1}(\mu_0)\), and then solves the unrestricted transparent-motives problem on the compressed domain \(K_L\). Equivalently, one can solve the unrestricted cheap-talk problem on the pseudo-state space \(\Omega_L^\ast:=\operatorname{Ext}(K_L)\).

\section{Generality of Belief-Threshold Games}\label{app:threshold-generality}

The baseline model of Section~\ref{sec:threshold} is the special case in which each risky action is tied to a singleton event. More generally, fix a default action $a_0$ and a risky action $a_i$, and define
\[
d_i(\omega):=u^R(a_i,\omega)-u^R(a_0,\omega).
\]
The next proposition characterizes exactly when the comparison between $a_i$ and the default has threshold form.

\begin{Proposition}\label{prop:event-threshold}
Fix a risky action $a_i$. The following are equivalent:
\begin{enumerate}[label=(\roman*)]
\item There exist an event $E_i\subseteq \Omega$ and a threshold $T_i\in(0,1)$ such that, for every posterior $\mu\in\Delta(\Omega)$,
\[
\sum_{\omega\in\Omega}\mu(\omega) d_i(\omega) \ge 0
\quad\Longleftrightarrow\quad
\mu(E_i)\ge T_i.
\]
\item There exist an event $E_i\subseteq \Omega$, a threshold $T_i\in(0,1)$, and a scalar $\lambda_i>0$ such that
\[
d_i(\omega)=
\begin{cases}
\lambda_i(1-T_i), & \omega\in E_i,\\
-\lambda_i T_i, & \omega\notin E_i.
\end{cases}
\]
Equivalently, after a positive affine normalization of payoff differences,
\[
d_i(\omega)=
\begin{cases}
\frac{1}{T_i}-1, & \omega\in E_i,\\
-1, & \omega\notin E_i.
\end{cases}
\]
\end{enumerate}
\end{Proposition}

\begin{proof}[\textbf{Proof of Proposition \ref{prop:event-threshold}.}]
If (ii) holds, then
\[
\sum_{\omega\in\Omega}\mu(\omega) d_i(\omega)
=
\lambda_i\bigl(\mu(E_i)-T_i\bigr),
\]
so the sign of the payoff difference is determined exactly by whether $\mu(E_i)$ exceeds $T_i$.

Now suppose (i) holds. Consider
\[
f_i(\mu)= \sum_{\omega\in\Omega}\mu(\omega) d_i(\omega),
\quad \text{ and } \quad
g_i(\mu) = \mu(E_i)-T_i.
\]
Both maps are affine on $\Delta(\Omega)$, they have the same zero set, and they have the same sign on each side of that hyperplane. Hence they must be proportional, so there exists $\lambda_i>0$ such that $f_i(\mu)=\lambda_i g_i(\mu)$ for every $\mu\in\Delta(\Omega).$ Evaluating this identity at posteriors concentrated on $\omega\in E_i$ and on $\omega\notin E_i$ yields (ii).
\end{proof}

When the events $E_1,\dots,E_m$ form a partition of $\Omega$, the original persuasion problem compresses exactly to a threshold game on the event space. Define the compressed state space $\Omega^*:=\{1,\dots,m\},$  the compressed prior $\mu_{0,i}^*:=\mu_0(E_i),$ and for each posterior $\mu$ define the compressed posterior $\mu^*_i:=\mu(E_i).$
Then action $a_i$ beats the default in the original problem if and only if $\mu^*_i\ge T_i$, and the sender's interim payoff depends on $\mu$ only through $\mu^*$. Thus the baseline singleton-state model is the special case $E_i=\{\omega_i\}$ and $T_i=T$.

Proposition~\ref{prop:event-threshold} shows that the singleton-state formulation in Section~\ref{sec:threshold} should not be read literally. In many applications, the primitive state space is finer than the receiver's decision problem: a risky action is justified not by one atomic state, but by a collection of underlying contingencies that all support the same action against the default. What matters for the threshold logic is therefore whether, for each risky action, beating the default is equivalent to the posterior probability of a single event crossing a cutoff.

When the action-justifying events \(E_1,\dots,E_m\) form a partition of the state space, the persuasion problem compresses exactly to a belief-threshold game on the induced event space. The results in Section~\ref{sec:threshold} and the heterogeneous-threshold extensions can then be read as applying to decision-relevant regimes or categories, rather than only to singleton states. This broadens the practical interpretation of the model to environments in which recommendations, certifications, diagnoses, or policy choices are triggered by sufficiently high posterior confidence in a coarse contingency.

The limitation is equally important. This is an exact reduction only for partition-event threshold problems. If the relevant events overlap, or if the comparison between a risky action and the default is not generated by a two-level payoff difference as in Proposition~\ref{prop:event-threshold}, then the baseline belief-threshold formulas need not apply.

\end{document}